\RequirePackage{lineno}
\documentclass[a4paper,11pt]{article}
\pdfoutput=1 

\usepackage{jheppub} 

\usepackage[T1]{fontenc} 

\input{defs}

\newcommand{\figref}[1]       {figure~\ref{fig:#1}}
\newcommand{\Figref}[1]       {Figure~\ref{fig:#1}}
\newcommand{\figsref}[2]      {figures~\ref{fig:#1} and \ref{fig:#2}}
\newcommand{\Figsref}[2]      {Figures~\ref{fig:#1} and \ref{fig:#2}}

\newcommand{\secref}[1]       {section~\ref{sec:#1}}
\newcommand{\secsref}[2]       {sections~\ref{sec:#1}~and~\ref{sec:#2}}

\newcommand{\appref}[1]       {appendix~\ref{app:#1}}

\newcommand{\appsref}[2]       {appendices~\ref{app:#1}~and~\ref{app:#2}}

\newcommand{\bibref}[1]       {ref.~\cite{#1}}

\usepackage{xspace}
\usepackage{graphicx}
\usepackage{subcaption}
\usepackage{relsize}

\title{\boldmath Pileup and Underlying Event Mitigation with Iterative Constituent Subtraction}

\author[a,1]{P. Berta,\note{Corresponding author.}}
\author[a]{L. Masetti,}
\author[b]{D. W. Miller,}
\author[c]{M. Spousta}

\affiliation[a]{PRISMA$^+$ Cluster of Excellence and Institute of Physics,  Johannes Gutenberg University Mainz,\\Staudingerweg 7, 55128 Mainz, Germany}
\affiliation[b]{The Enrico Fermi Institute and the Department of Physics, University of Chicago,\\5640 S. Ellis Ave, Chicago, IL, 60637, USA}
\affiliation[c]{Institute of Particle and Nuclear Physics, Faculty of Mathematics and Physics, Charles University, V Hole\v sovi\v ck\' ach 2, 180 00 Prague 8, Czech Republic}

\emailAdd{peberta@uni-mainz.de}
\emailAdd{masetti@uni-mainz.de}
\emailAdd{David.W.Miller@uchicago.edu}
\emailAdd{Martin.Spousta@mff.cuni.cz}

\abstract{

The hard-scatter processes in hadronic collisions are often largely contaminated with soft background coming from pileup in proton-proton collisions, or underlying event in heavy-ion collisions. This paper presents a new background subtraction method for jets and event observables (such as missing transverse energy) which is based on the previously published Constituent Subtraction algorithm. The new subtraction method, called Iterative Constituent Subtraction, applies event-wide implementation of Constituent Subtraction iteratively in order to fully equilibrate the background subtraction across the entire event. Besides documenting the new method, we provide guidelines for setting the free parameters of the subtraction algorithm. Using particle-level simulation, we provide a comparison of Iterative Constituent Subtraction with several existing methods from which we conclude that the new method has a significant potential to improve the background mitigation in both proton-proton and heavy-ion collisions.

}

\begin{document} 

\maketitle
\flushbottom

\section{Introduction}
\label{sec:intro}

  Precision tests of Standard Model of particle physics as well as searches for new physics at the Large Hadron Collider (LHC) require maximizing the collected data 
which is primarily achieved by increasing the beam intensities. The dilemma is that high intensities also increase the 
number of simultaneous proton-proton ($pp$) interactions in a single colliding bunch crossing of the two beams. Such multiple $pp$ collisions, also known as 
\textit{\pileup}, are then read out as one single event. 
  Pileup contributes a primarily low transverse momentum (\pt) and largely diffuse background of particles to the hard scattering process of interest that subsequently distorts many hadronic final state observables, in particular the kinematics and substructures of hadronic jets.

In the last part of LHC Run~2, the ATLAS and CMS experiments achieved an average and peak pileup of ${\sim}35$ and ${\sim}70$ $pp$ simultaneous collisions, 
respectively~\cite{ref:pileup1,ref:pileup2}.
Each additional \pileup $pp$ collision at $13\tev$ adds an average \pt of approximately ${\sim}900\mev$ per unit area in the rapidity-azimuth (\yphi) plane. In the upcoming LHC Run~3, an average \pileup of 70 collisions per bunch crossing is expected, corresponding to the maximum \pileup level of Run~2. The high-luminosity LHC is expected to deliver an average pileup of 200 $pp$ collisions per bunch crossing~\cite{Atlas:2019qfx}.  An even more intense environment is present in 
heavy-ion collisions at the LHC and at the Relativistic Heavy Ion Collider (RHIC) where a large underlying event (UE) may lead to a background \pt of more than $300\gev$ per unit area in \yphi space~\cite{Chatrchyan:2011sx}.
Efficient techniques need to be developed and tested in order to mitigate the impact of these large 
backgrounds on jet kinematics, jet substructure, and missing \et measurements. Throughout this paper the word \textit{background} will be used to refer either to \pileup 
in $pp$ collisions or to the UE in heavy-ion collisions.

One category of background mitigation techniques includes subtraction methods or algorithms that do not alter the jet definition itself. Examples of algorithms in this 
category include the extensively used \areaSubtraction~\cite{Cacciari:2007fd,Cacciari:2008gn} approach to correct jet kinematics, and its extension to corrections for 
jet shape observables, the \textit{Shape-expansion} method~\cite{Soyez:2012hv}. Also in this category are numerous algorithms aimed at correcting the inputs to jet 
reconstruction prior to the application of a jet algorithm: Constituent Subtraction (CS) \cite{Berta:2014eza}, SoftKiller~\cite{Cacciari:2014gra}, 
PUPPI~\cite{Bertolini:2014bba}, and jet cleansing~\cite{Krohn:2013lba} are among the most widely studied. A second category of background mitigation techniques consists 
of the so-called \textit{jet grooming} methods, which are often used in the context of highly Lorentz-boosted massive objects, to both improve the precision of the 
reconstruction itself and mitigate the effects of pileup and UE. Grooming algorithms fundamentally modify the jet definition to be less sensitive to these backgrounds. 
The most frequently used methods are filtering~\cite{Butterworth:2008iy}, trimming~\cite{Krohn:2009th}, pruning~\cite{Ellis:2009me}, and soft drop methods
\cite{Larkoski:2014wba,Dasgupta:2013ihk,Dreyer:2018tjj}, each of which have configurable parameters that determine their properties as 
\textit{groomers}.
 These methods often work with subjets that are determined by
 the \ca algorithm~\cite{Dokshitzer:1997in,Wobisch:1998wt} or the \kt~algorithm \cite{Catani:1993hr,Ellis:1993tq} applied to a large-radius \antikt~\cite{Cacciari:2008gp} jet. 
Specific algorithms or criteria are then applied to these subjets to eliminate soft and/or wide-angle contributions to a jet, which are the most likely to be contaminated by backgrounds. 
  Besides subtraction and grooming methods, techniques based on machine learning have also been implemented and tested recently~\cite{Komiske:2017ubm,Martinez:2018fwc,Carrazza:2019efs}. 
  The potential of improving background mitigation by incorporating charged-track information to subtract neutral \pileup is discussed in \bibref{Cacciari:2014jta}. 
  Recently, a novel approach to estimate the background is proposed in \bibref{Yacine:2019ycj}, which aims at reducing the impact of fluctuations 
in the background on the jet observables and which may be potentially combined with CS algorithm or other subtraction method.
  A detailed overview of background mitigation techniques can be found in \bibref{Soyez:2018opl}.

  This paper presents a new background subtraction method for jets, jet substructure observables, and global event observables such as missing \et, which is based on 
the CS method.
  In contrast to other background subtraction methods, such as the \areaSubtraction or the shape-expansion method, CS corrects for background at the 
level of jet constituents, which may be particles, tracks from an inner detector, or calorimeter clusters~\cite{Bellettini:1982gi,Lampl:2008zz,Sirunyan:2017ulk}). Jet kinematics and substructure are simultaneously corrected with this approach.
  The CS method was successfully used in several measurements at the LHC. The ALICE and CMS experiments use CS in measurements of jet shapes and mass in heavy-ion 
collisions~\cite{Acharya:2018uvf,Acharya:2017goa,Sirunyan:2018gct}, furthermore CMS used CS in the measurement of splitting functions~\cite{Sirunyan:2017bsd}. 
  The CS method has been applied in several performance studies by CMS~\cite{CERN-CMS-DP-2018-024,CMS-PAS-JME-14-001} and 
ATLAS~\cite{ATL-PHYS-PUB-2017-020,ATL-PHYS-PUB-2018-011,ATLAS:2017pfq} and it is being discussed and tested in the context of future experiments, in particular the 
Compact Linear Collider (CLIC) and Future Circular Collider (FCC)~\cite{Boronat:2016tgd,Golling:2016gvc}.
  The CS method was also used in a recent measurement of jet substructure by STAR at RHIC~\cite{KunnawalkamElayavalli:2019csg}.
  Phenomenological 
studies also consider the CS method, in particular in the context of tagging boosted bosons and top quarks~\cite{Behr:2015oqq,Salam:2016yht,Larkoski:2015yqa}, searches 
for new physics \cite{Berlin:2015aba,Craig:2014lda,Brust:2014gia,Low:2014cba}, and structure of parton shower~\cite{Dreyer:2018nbf}. The new method for background 
subtraction presented in this paper may help to improve the precision of measurements of Standard Model processes and may make experimental studies less susceptible to increasing 
backgrounds at colliders.  This new method is applicable for the subtraction of both \pileup in $pp$ collisions and UE in heavy-ion collisions. In this paper, we test the methods explicitly only in the environment of $pp$ collisions with \pileup. 

This paper is organized as follows. First, in \secref{algo}, an event-wide background subtraction performed using CS is discussed, which was briefly mentioned in \bibref{Berta:2014eza} as a possible extension of the original CS method. In \secref{ics}, the new method for background subtraction is introduced. Then, in \secref{performance}, performance of various methods for background subtraction including the new method is presented in the context of jet reconstruction and substructure observables.

\section{Event-wide pileup mitigation with CS}
\label{sec:algo}
The CS algorithm described in \bibref{Berta:2014eza} corrects individual jets which were already clustered using a certain jet algorithm. We refer to 
this approach as \textit{Jet-by-jet CS}. However, the jet clustering can be biased by the presence of background resulting in different particle content of jets (this is referred to as ``back reaction'' in \bibref{Cacciari:2008gn}). As briefly mentioned in \bibref{Berta:2014eza}, the CS algorithm can be extended to correct the event constituents before jet clustering. The jets resulting from this \textit{Event-wide CS} correction followed by jet clustering can be more precise than the individually corrected jets. In the following, the Event-wide CS is described assuming massless inputs. A discussion of the correction of massive inputs can be found in 
\appref{massive_inputs}.

The basic ingredient of CS is the background \pt density, $\rho$, which was introduced in the \areaSubtraction. Several methods to estimate this quantity are described in \cite{Cacciari:2011ma}. In general, $\rho$ can be estimated as a function of other variables, most commonly as a function of rapidity. The estimated $\rho$ is then used to scale the \pt of the \textit{ghosts} in the Event-wide CS. The ghosts are infinitly soft particles (in practice $\pt\approx10^{-100}\gev$) incorporated into the event such that they uniformly cover the \yphi plane with high density. Each of these ghosts is massless and covers a fixed area, \Aghost, in the \yphi plane. Historically, the ghosts can be used to define the jet area \cite{Cacciari:2007fd,Cacciari:2008gn} for the \areaSubtraction method or perform background subtraction in the Shape-expansion and Jet-by-jet CS methods. In all these methods, their property of infinite softness is essential to not modify the jet clustering sequence. However, for the Event-wide CS, this property is irrevelant and each ghost \pt is directly set to $\ghostpt\equiv\Aghost\cdot\rho$. Then such ghosts already represent the expected background contribution in the given event, and can be used to correct particles via the Event-wide CS as follows.

Only particles and ghosts with pseudo-rapidity, $\eta$, fulfilling $|\eta|<\maxEta$ are used in the correction procedure. The parameter \maxEta defines the detector acceptance of particles.\footnote{In the software implementation of the Event-wide CS algorithm in \texttt{FastJet Contrib} \cite{Salam:2019:fastjetContrib}, the user can define arbitrary phase-space using the \texttt{fastjet::Selector} class from \texttt{FastJet}~\cite{Cacciari:2011ma, Cacciari:2005hq}. In that way, phase space asymmetric in $\eta$ can be used or the correction can be done only for particles below certain \pt threshold.} For each pair of particle $i$ and ghost $k$, a matching scheme is implemented using the distance measure, $\Dik$, defined as
  %
\begin{equation}
  \Dik=\pti^\alpha \cdot \DeltaRik,
  \label{eq:Dik}
\end{equation}
  %
where $\alpha$ is a free parameter and \DeltaRik is defined as
  %
\begin{equation}
\DeltaRik=\sqrt{\left(\yi-\ghostyk\right)^2+\left(\phii-\ghostphik\right)^2}.
  \label{eq:deltaRik}
\end{equation}
  %
The list of all distance measures, $\{\Dik\}$, is sorted from the lowest to the highest values. The background removal proceeds iteratively, starting from the particle-ghost pair with the lowest $\Dik$. At each step, the momentum \pt of each particle $i$ and ghost $k$ are modified as follows.
  %
\begin{equation}
\begin{split}
\text{If}~ \pti\geq\ghostptk: ~~~~~
   &\pti\longrightarrow\pti-\ghostptk, \\
   & \ghostptk\longrightarrow 0; \\
\text{otherwise:}~~~~~~
   & \pti\longrightarrow 0, \\
   & \ghostptk\longrightarrow \ghostptk-\pti. \\
\end{split}
  \label{eq:correction}
\end{equation}
  %
The iterative process is terminated when $\DeltaRik > \DeltaRmax$ where \DeltaRmax is a free parameter.\footnote{The meaning of the \DeltaRmax parameter was different in the original description of the CS procedure in \bibref{Berta:2014eza}. In the software implementation in \texttt{FastJet Contrib}, the meaning described in this paper is used since version 1.022.} The output of the correction procedure is a set of \fourmomenta representing the background-corrected event. Any operation can be done on these output particles - most commonly jet clustering, evaluation of global event shapes or missing transverse energy. Besides providing the ability to correct the full event, which is documented for the case of missing transverse energy in \secref{missinget}, the performance of the background subtraction is also improved with respect to the original CS as documented in \secref{performance}. A detailed discussion of the choice of the free CS parameters is provided in \appref{CS_parameters}.

\section{Iterative Constituent Subtraction}
\label{sec:ics}
\begin{figure}[!h]
  \centering
  \begin{subfigure}{0.46\textwidth}
    \includegraphics[width=\textwidth]{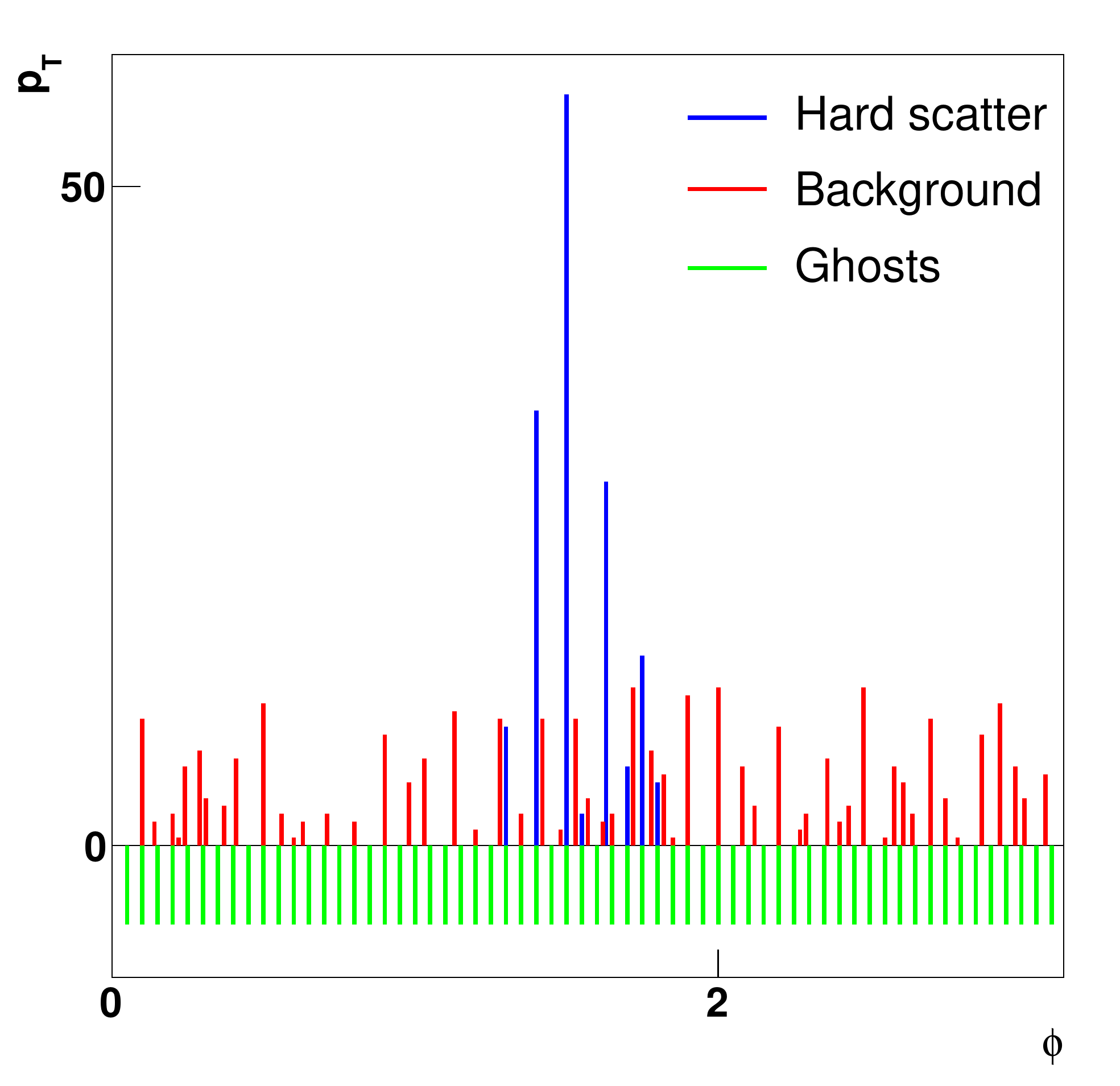}
    \caption{}
    \label{fig:before_subtraction1}
  \end{subfigure}
  \begin{subfigure}{0.46\textwidth}
    \includegraphics[width=\textwidth]{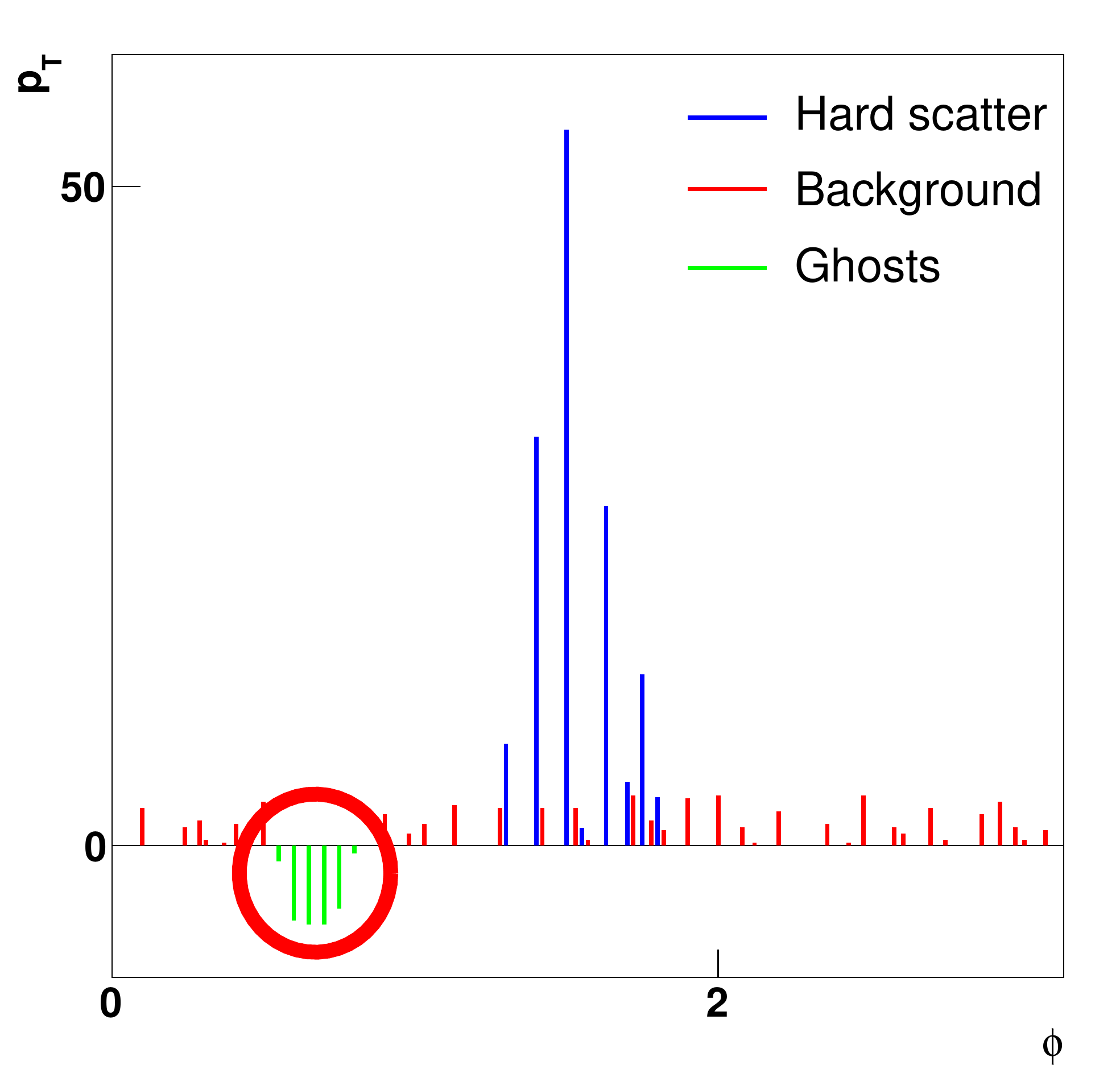}
    \caption{}
    \label{fig:after_subtraction1}
  \end{subfigure}
\\
\begin{subfigure}{0.46\textwidth}
    \includegraphics[width=\textwidth]{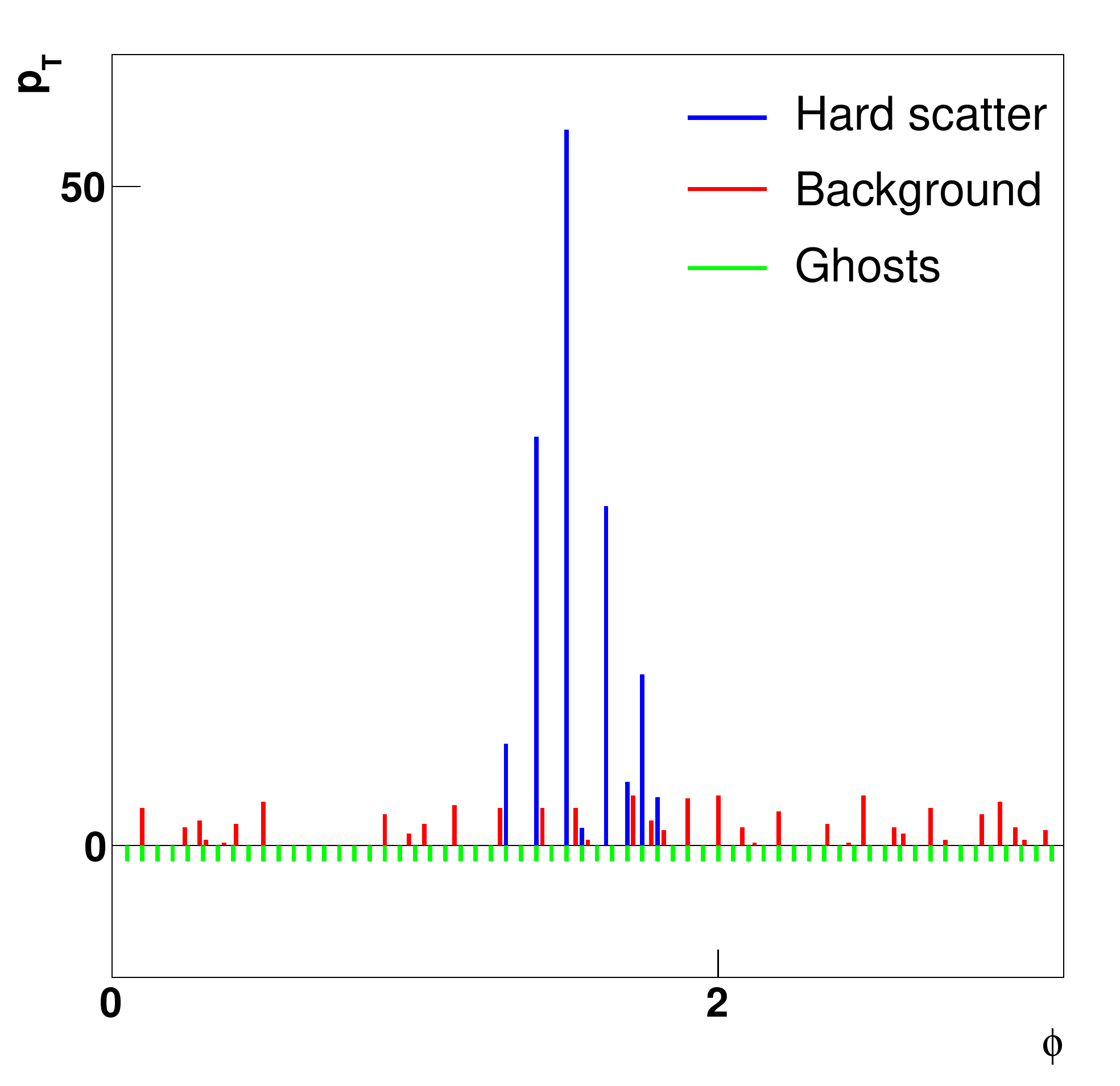}
    \caption{}
    \label{fig:before_subtraction2}
  \end{subfigure}
  \begin{subfigure}{0.46\textwidth}
    \includegraphics[width=\textwidth]{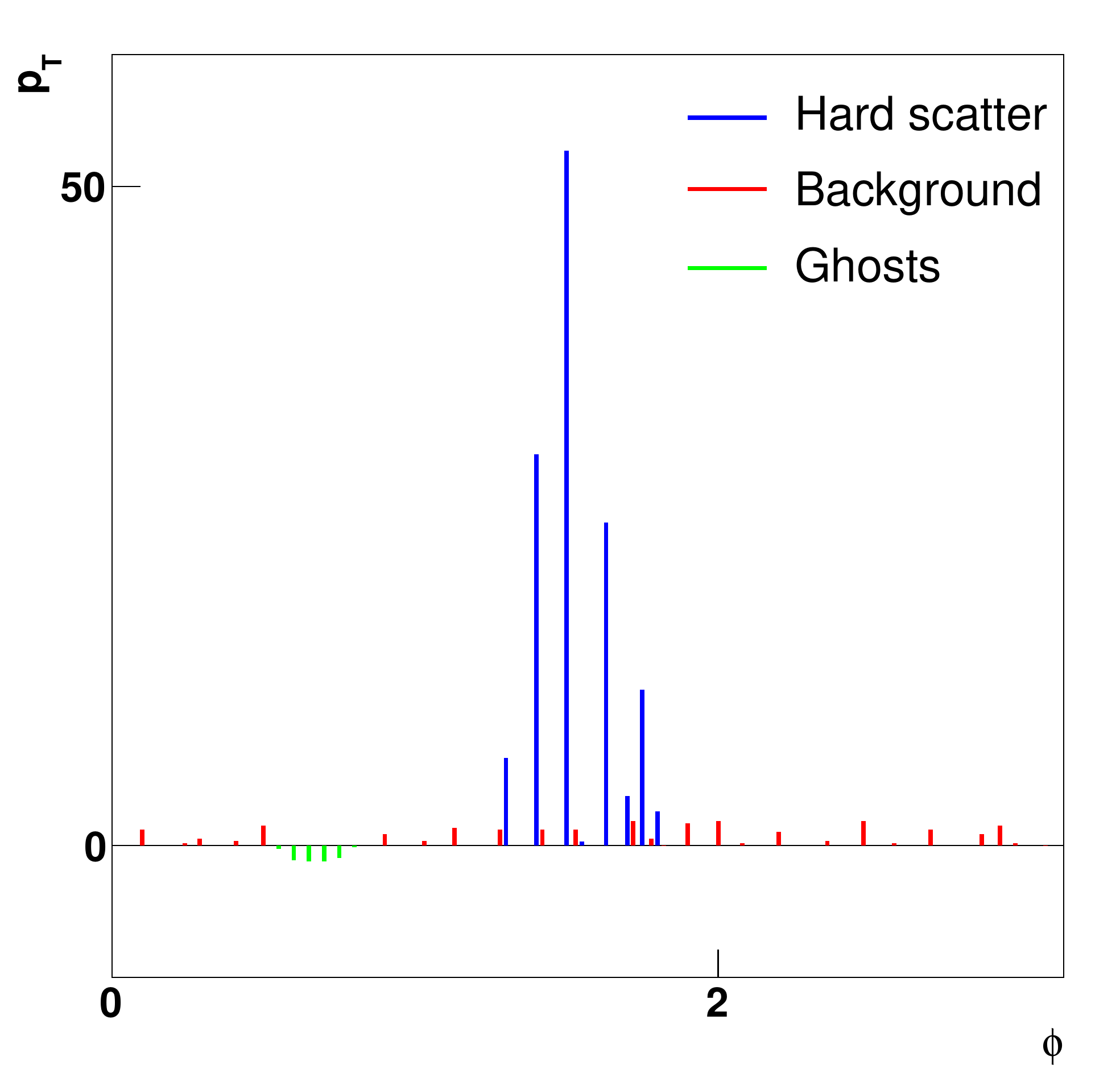}
    \caption{}
    \label{fig:after_subtraction2}
  \end{subfigure}
  \caption{
  Illustration of the ICS method with $2$ iterations assuming only one dimension in azimuth. The event before the first iteration (\ref{fig:before_subtraction1}) 
contains hard-scatter particles, background particles and ghosts with \pt corresponding to $\rho$ in that event. To emphasize the fact that the ghosts are subtracted, 
their \pt is negative in these figures.  After the first iteration with finite \DeltaRmax (\ref{fig:after_subtraction1}), ghosts with unsubtracted \pt\ can remain in 
the event (emphasized by a red circle). The scalar \pt sum of the unsubtracted ghosts is redistributed uniformly in the \yphi plane (\ref{fig:before_subtraction2}) or 
according to the actual $\rho$ dependence. After the second iteration (\ref{fig:after_subtraction2}), the unsubtracted background \pt is strongly reduced.
  }
  \label{fig:ICS_illustration}
\end{figure}

Iterative Constituent Subtraction (ICS) is a new background mitigation method that extends the concept of the original CS method. ICS applies the event-wide implementation of CS iteratively with finite \DeltaRmax value. After each iteration, any remaining unsubtracted background estimate is redistributed uniformly across the entire event and another CS procedure is performed. The exact algorithmic procedure using \Niter iterations is the following:

\begin{enumerate}
\item Estimate $\rho$ using a given estimation method and create ghosts with $\ghostpt=\rho\Aghost$. The $\rho$ can be a function of other variables.

\item Perform Event-wide CS correction using a finite $\DeltaRmax$ value. Define \textit{input ghosts} as ghosts present in the event before this CS correction. Define \textit{output ghosts} as ghosts remaining in the event after this CS correction. 

\item Compute the scalar \pt sum of the input ghosts, \ptInput, and of the output ghosts, \ptOutput. Update the input ghosts by scaling their \pt by a factor $\ptOutput/\ptInput$.

\item Perform next iteration by going back to step 2 using the updated input ghosts. The parameters of CS, such as \DeltaRmax, may be changed for each iteration.

\end{enumerate}
  The step 4 is performed ($\Niter-1$)-times. The algorithm with $\Niter=2$ is illustrated on an example event in \figref{ICS_illustration}.

An option exists for the above algorithm that avoids the usage of ghosts in the second or higher iteration which were not fully subtracted in the previous iteration. 
  With this option, the \ptInput is evaluated using only ghosts which were fully subtracted in the previous iteration, while the unsubtracted ghosts are discarded for the actual iteration. This approach avoids placing the expected background 
deposition to positions where there is small chance for combining ghosts with real particles in the second or higher iteration. We refer to this option as \textit{ghost removal}. The impact of this option on the performance depends heavily on the chosen \DeltaRmax parameters for each iteration.

In our performance studies, it was found that ICS outperforms both Jet-by-jet CS and Event-wide CS (described in \secref{algo}) as documented in \secref{performance}. A detailed discussion of the choice of the configuration of the ICS method is provided in \appref{ICS_parameters}.
The ICS method is implemented in \texttt{FastJet Contrib} \cite{Salam:2019:fastjetContrib} since version 1.038.

\section{Performance}
\label{sec:performance}
In this section, we discuss the performance of new methods presented in \secsref{algo}{ics}. The definitions of jet shapes which are commonly used in jet substructure 
studies are summarized in \secref{definitions}. In \secref{quantifying}, the metrics used to measure the performance of background subtraction methods is discussed. The samples 
and general settings used for the performance evaluation are described in \secref{performance_setup}. 
  The other sections contain several performance studies. Studies for ICS, Event-wide CS, Jet-by-jet CS and \areaSubtraction for jets are presented in \secref{jet}. 
Studies for missing transverse energy are presented in \secref{missinget}. An extensive comparison among a several selected methods using a common open-source software 
framework for background subtraction performance comparisons is provided in \secref{performance_workshop}.

\subsection{Jet shape definitions}
\label{sec:definitions}
While kinematic properties of jets can be uniquely characterized by four-momentum, $P^\mu = (p_\mathrm{T}, \eta, \phi, m)$, the internal structure of jet cannot be 
characterized by a single observable. Instead, various observables have been proposed in order to capture many internal properties of jets. These different observables are 
generically referred to as jet shapes. This term extends the meaning of the observable $\rho(r)$, originally called jet shape, which was measured at various experiments 
\cite{Akers:1994wj,Abachi:1995zw,Adloff:1998ni,Chekanov:2004kz,Acosta:2005ix,Aad:2011kq,Chatrchyan:2012mec} and which quantifies the radial energy flow, from the jet axis 
$r = \sqrt{(\Delta\eta)^2 + (\Delta\phi)^2}$. This quantity, along with other observables such as fragmentation function, can be used to quantify the 
internal structure of QCD jets in the context of the parton shower evolution. More recently, jet shape observables have been extensively employed as tools in 
measurements of Lorentz-boosted massive objects \cite{Abazov:2011vh,Aaltonen:2011pg,Khachatryan:2014vla,Aad:2013gja}. In such cases, jets have fundamentally different internal 
structures as compared to jets formed from light quarks and gluons. One canonical example is the two-prong structure formed from a collimated pair of heavy quarks from the decay of a boosted Higgs boson 
\cite{Butterworth:2008iy}. Various jet shape and structure observables can be used to distinguish jets formed by massive boosted object decays from jets formed solely by parton showering and hadronization of light quarks and gluons. In this 
study, we use three representative jet shape observables -- \textit{jet width}, two-to-one and three-to-two ratios of \textit{N-subjettiness} 
($\tau_{21}$ and $\tau_{32}$) -- and the \textit{jet mass}.

Jet mass, $m = \sqrt{P^\mu P_\mu}$, is a basic observable used to identify boosted hadronically decaying objects such as the $W^{\pm}$~boson (see e.g. \bibref{Aad:2015rpa}).
Jet width (also known as \textit{linear radial moment} or \textit{girth}) is defined as the first moment of the radial flow of transverse momentum or energy,
\begin{equation}
\mathrm{jet~width} = \frac{\sum_i  \pti \Delta R_i}{\sum_i \pti},
\end{equation}
where \pti is magnitude of the transverse momentum of jet constituent $i$ and $\Delta R_i$ is the distance between jet constituent $i$ and the jet axis.
  Jet width has been shown to provide e.g. a good discriminating power between quark-initiated and gluon-initiated jets \cite{Gallicchio:2011xq}.

  The $\tau_{21}$ and $\tau_{32}$ are two-to-one and three-to-two ratio of $N$-subjettiness, respectively. 
Subjettiness was introduced to provide an enhanced discriminating power for identifying boosted massive objects 
\cite{Thaler:2010tr,Thaler:2011gf}.
The $N$-subjettiness is defined as
  \begin{equation}
  \tauN = \frac{1}{d_{0}} \sum_k p_{\mathrm{T}k} \cdot  \text{min}(\Delta R_{1k}, \Delta R_{2k},...,\Delta R_{Nk})~\text{,~~with},
  ~~~d_{0}\equiv\sum_{k} p_{\mathrm{T}k}\cdot  R
  \label{eq:nsubj}
  \end{equation}
  %
  where $R$ is the distance parameter of the jet algorithm, $p_{\mathrm{T}k}$ is the transverse momentum of constituent~$k$ and $\Delta R_{ik}$ is the distance between 
a subjet~$i$ and a constituent~$k$. The $N$ subjets are defined by re-clustering the constituents of the jet with exclusive version of the \kt algorithm and requiring 
that exactly $N$ subjets are found.

\subsection{Quantifying performance of background subtraction}
\label{sec:quantifying}
The following figures of merit are used to quantify the performance of jet reconstruction~\cite{Chatrchyan:2011ds,Aad:2011he}: jet energy scale (JES), jet energy resolution (JER), 
jet reconstruction efficiency, and rate of fake jets. JES is also sometimes called linearity. These quantities are calculated using Monte Carlo generators coupled with full detector simulations for a particular experiment by comparing particle-level jets (so-called \textit{true jets}) with jets reconstructed using the outputs of the detector simulation. JES and JER characterize the mean and 
root-mean-square, respectively, of the difference between the transverse momentum of particle level jet, $\pttrue$, and the transverse momentum of jet 
reconstructed in the detector, $\ptreco$. More explicitly, $\mathrm{JES} = \avg{\ptreco / \pttrue}$ and $\mathrm{JER} = \sigma( \ptreco / \pttrue )$. The JES is determined by a combination of the
 response of the detector and the performance of algorithms used for the jet calibration and background subtraction. The JER of calorimeter 
jets can be factorized as follows
  \begin{equation}
  \sigma\bigg(
     \frac{\ptreco} {\pttrue}
  \bigg) = \frac{a}{\sqrt{ \pttrue}}\oplus\frac{b}{ \pttrue}\oplus{c},
  \label{eq:jer}
  \end{equation}
  where $a$ is the stochastic term, $b$ is the noise term and $c$ is the constant term \cite{Aaboud:2018twu,Fabjan:2003aq}. The stochastic and constant terms are governed 
by the response of the detector to the particle shower. The noise term is largely determined by fluctuations of backgrounds, which may include electronic noise, pileup, 
or the underlying event. At the LHC, the contributions due to pileup tend to dominate. In the case of an average background subtraction such as the \areaSubtraction, the noise term is given by the root-mean-square of \pt 
evaluated in the area of a jet excluding the jet signal, $b = \mathrm{RMS}(p_\mathrm{T}^{\mathrm{area}})$. The stochastic and constant terms are not significantly affected by the 
subtraction algorithm~\cite{Aaboud:2018twu}, but they can be reduced by using the information about the jet internal structure in the calibration procedure~\cite{Aad:2012ag} or by combining the information from calorimeter with the information from tracking~\cite{CMS:2009nxa}. The noise term cannot be reduced by the 
calibration procedure, but it can be reduced by subtraction procedure or using some additional information about 
the background compared to the basic information about its average density. This can be, for example, the information about pointing of jet constituents to the primary 
vertex (e.g. \textit{charged hadron subtraction} used by CMS~\cite{CMS:2014ata} or \textit{jet-vertex association} used by 
ATLAS~\cite{ATL-PHYS-PUB-2014-001}) or by noise suppression at the sub-constituent level of calorimeter cells~\cite{Abazov:2013hda,Aad:2016upy}.

Jet reconstruction efficiency quantifies the probability that a jet is found given the presence of a true jet, subject to specific kinematic constraints for a particular efficiency 
evaluation. As such, the jet reconstruction efficiency is indirectly affected by both the JES and the JER. The rate of fake jets is important in the case of large backgrounds, 
such as in heavy-ion collisions, where correlated background fluctuations can lead to a large rate of fake jets. Consequently, this rate 
can be reduced by reducing the fluctuations in the background for which the JER is the relevant  performance metric. We therefore focus on the JES and JER in order to characterize the jet reconstruction performance.

The JES and JER can be generalized to any of the observable quantities that characterize the jet kinematics or shapes discussed in \secref{definitions}. We define the \textit{bias} and \textit{resolution} of a given observable, $x$, as follows:

  \begin{eqnarray}
  \mathrm{bias} = \frac{\langle x^{\mathrm{rec}} - x^{\mathrm{true}}\rangle}{\langle x^{\mathrm{true}}\rangle}, ~~~
  \mathrm{resolution} = \frac{\mathrm{RMS}( x^{\mathrm{rec}} - x^{\mathrm{true}})}{\langle x^{\mathrm{true}}\rangle}.
  \end{eqnarray}

The difference between $x^{\mathrm{rec}}$ and $x^{\mathrm{true}}$ is used instead of their ratio in order to avoid excessive values of bias for small values of 
$x^{\mathrm{true}}$. The denominator in the bias and resolution allows for an easier comparison among different quantities.
  Similarly to the case of the JES and JER, these quantities characterize the primary aspects of the performance of the background subtraction.
These quantities are therefore evaluated simultaneously for all the jet shape observables discussed in \secref{definitions}.

\subsection{Test samples and configuration of subtraction}
\label{sec:performance_setup}
The performance of subtraction methods is evaluated in simulated $pp$ collisions at $\sqrt{13}\tev$ using events with boosted top quarks from the decay $Z' \rightarrow t\bar{t}$ of a hypothetical boson $Z'$ with a mass of 
$1.5\tev$. These simulated hard-process events are referred to as \textit{true} events. To obtain the \textit{reconstructed} events with \pileup included, the true events are overlaid with inclusive  $pp$ collision events that represent \pileup. The number of overlaid inclusive events, \npu, has a uniform distribution from 0 to 140.
All event generation is performed with PYTHIA 8.180~\cite{Sjostrand:2006za,Sjostrand:2007gs} using the tune 4C and CTEQ 5L parton density functions~\cite{Lai:1999wy}. The hard process is generated without underlying event.

A pseudo-detector simulation is then used. All particles are grouped into towers of size $0.1 \times 0.1$ in the pseudorapidity-azimuth (\etaphi) space. The tower energy is 
obtained as the sum of energies of particles pointing to that tower. All neutrinos and muons are discarded. Only towers with $|\eta|<4.0$ are selected. The mass of each 
tower is set to 0. The $\eta$ and $\phi$ of each tower is randomly smeared using a Gaussian kernel with width of 0.1 (but maximally up to 0.2).

  The reconstructed events are corrected with both ICS and multiple other \pileup mitigation techniques. To evaluate the performance of the various methods, the jets from the corrected 
events are compared to jets from the true events using the quantities defined in \secref{quantifying}. To factorize the effect of \pileup, the same detector 
simulation is used for both, true and reconstructed, events. Two jet definitions are used: \akt algorithm with the distance parameter $0.4$ and $1.0$. 
  Only true jets with $|\eta|<3$ and $\pt>20\gev$ are used.

All jet finding and background estimation is performed using \texttt{FastJet} 3.3.1~\cite{Cacciari:2011ma, Cacciari:2005hq}. The event energy density $\rho$ is 
estimated as a function of $y$ using the \texttt{Grid\-Median\-Background\-Estimator} tool from \texttt{FastJet} with a grid spacing of 0.5 using the particles up to $|y|= 4$. Since the inputs are massless, there is no need to derive the background estimate for the mass term of \pileup, \rhom. The same $\rho$ estimation is used for the \areaSubtraction and the CS-based methods. The \areaSubtraction method was carried out as 4-vector subtraction using the \texttt{fastjet::Subtractor} class from \texttt{FastJet} 3.3.1. Unphysical situations with negative corrected mass are avoided by enabling the \texttt{safe\string_mass} option.

For the CS-based methods, we used \texttt{FastJet Contrib} version 1.042. The choice of the optimal CS parameters and ICS configurations is discussed in detail in \appsref{CS_parameters}{ICS_parameters}. The following configurations are found to be optimal for the majority of studies presented in \secsref{jet}{missinget}. The Jet-by-jet CS uses parameters $\DeltaRmax=\infty$, $\alpha=0$, and $\Aghost=0.0025$ for both jet definitions. The Event-wide CS uses parameters $\alpha=1$, $\Aghost=0.0025$, and \DeltaRmax parameters $\DeltaRmax=0.25$ and $\DeltaRmax=0.7$ for \aktfour and $R=1.0$ jet definitions, respectively. The ICS method is used with two iterations using the ghost removal option. For both iterations, the same $\alpha$ and \Aghost parameters are used, $\alpha=1$ and $\Aghost=0.0025$, while the \DeltaRmax is different for each iteration and also depends on the jet definition. For \aktfour jets, parameters $\DeltaRmax_1=0.2$ and $\DeltaRmax_2=0.1$ are used for the first and second iteration, respectively. For \aktten jets, the parameters $\DeltaRmax_1=0.2$ and $\DeltaRmax_2=0.35$ are used.

\subsection{Performance for jet kinematics and substructure}
\label{sec:jet}
The performance of the ICS method applied to jets is evaluated for both jet kinematic and jet shape observables, which are defined in \secref{definitions}. The bias and 
resolution (defined in \secref{quantifying}) for the observables are studied as a function of number of \pileup interactions (\npu) and jet \pt, as well as for two 
choices of \akt distance parameter, $R=0.4$ and $R=1.0$. The performance of ICS is compared to the performance of the \areaSubtraction, Jet-by-jet CS (both 
introduced in \secref{intro}), and the Event-wide CS (discussed in \secref{algo}) using the configurations described in \secref{performance_setup}. 

\begin{figure}[!h]
  \centering
  \begin{subfigure}{0.47\textwidth}
    \includegraphics[width=\textwidth]{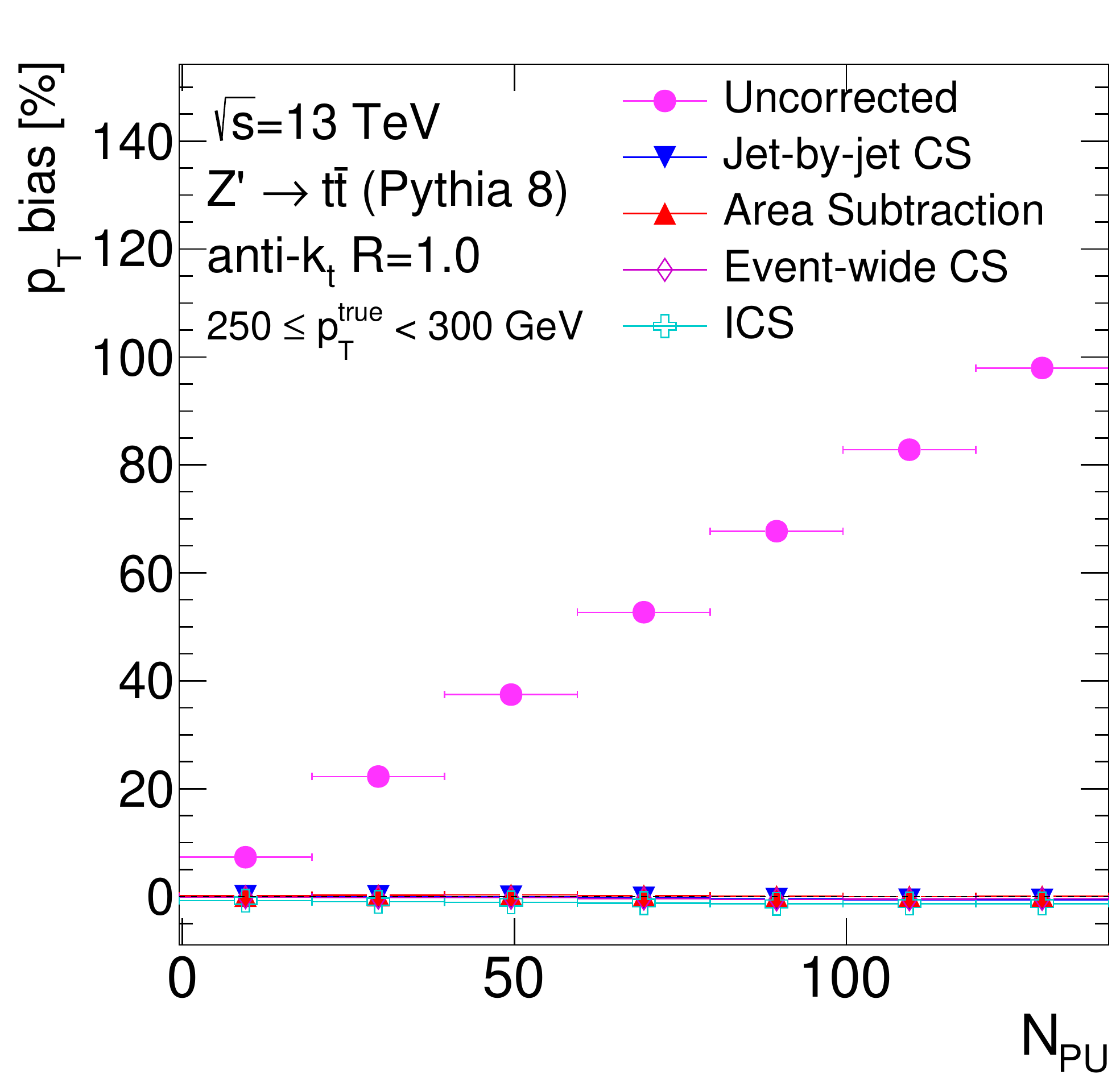}
    \caption{}
    \label{fig:Zprime:pt_bias}
  \end{subfigure}
  \begin{subfigure}{0.47\textwidth}
    \includegraphics[width=\textwidth]{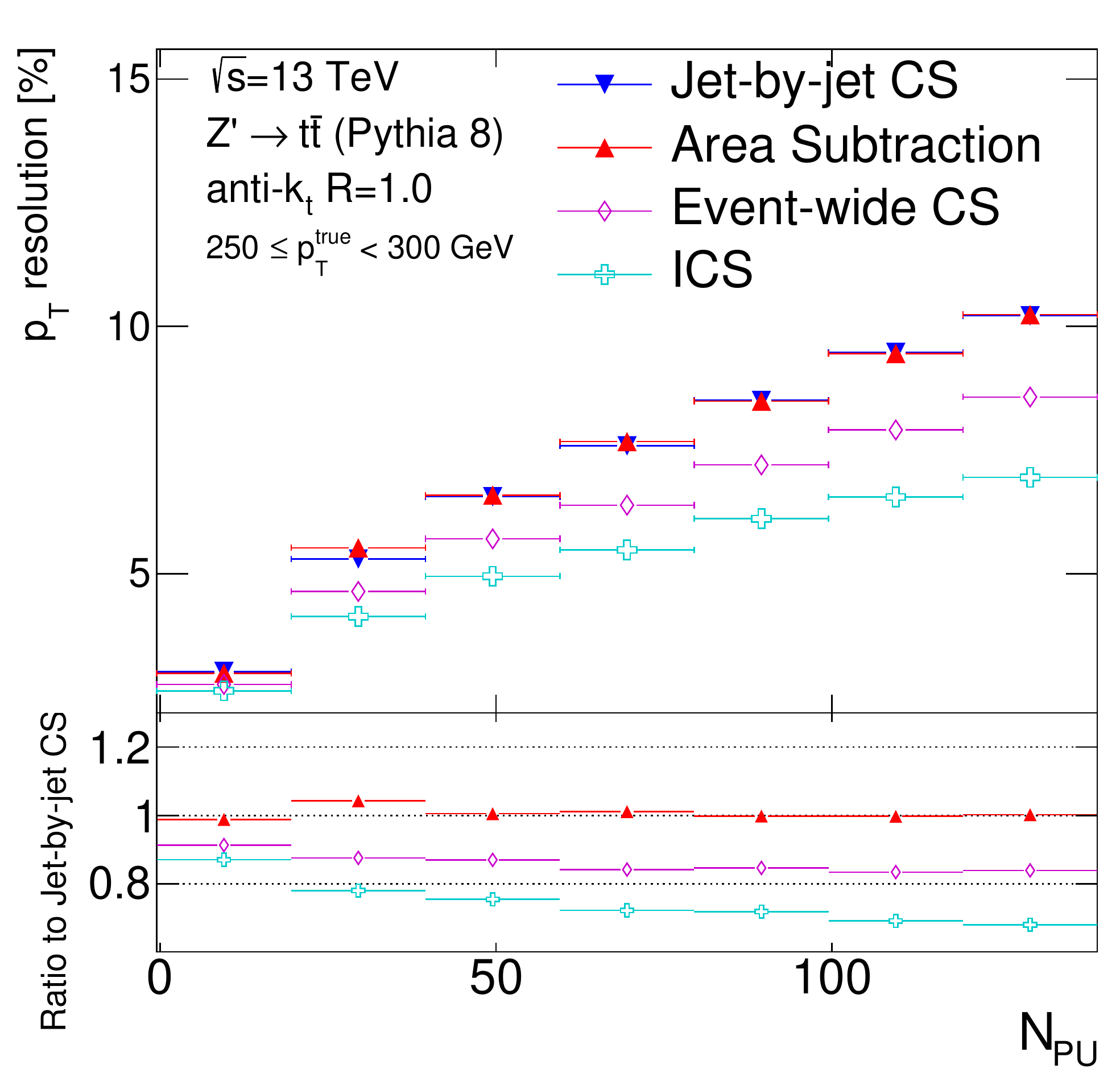}
    \caption{}
    \label{fig:Zprime:pt_resolution}
  \end{subfigure}
  \caption{Dependence of jet \pt bias (left) and resolution (right) on \npu for four \pileup correction methods (Jet-by-jet CS, \areaSubtraction, Event-wide CS, 
and ICS).}  
  \label{fig:Zprime:pt}
\end{figure}

Perhaps the most illustrative demonstration of the efficiency of any \pileup correction is the extent to which a given algorithm is able to reduce the dependence of an 
observable on the amount of \pileup in the event, as parameterized by \npu. \Figsref{Zprime:pt}{Zprime:mass} show the impact of \pileup on the \pt and mass, 
respectively, for large-radius ($R=1.0$) jets containing the decay products of boosted top-quarks from $Z' \rightarrow t\bar{t}$ (see \secref{performance_setup}). 
Only a narrow true jet \pttrue\ range is chosen for these figures, $250\gev\leq \pttrue < 300\gev$. The four correction algorithms considered for comparison demonstrate 
the improvements achievable in terms of the bias (\figsref{Zprime:pt_bias}{Zprime:mass_bias}) and resolution (\figsref{Zprime:pt_resolution}{Zprime:mass_resolution}) in each case, as a function of \npu.

Each of the algorithms considered is able to remove the bias introduced by \pileup in both the \pt\ and the mass of these jets to approximately the same degree. 
However, the precision of these corrections, as quantified by the resolution of the corrected measure, can vary significantly. The Event-wide CS and the ICS corrections 
are both observed to improve the resolution of the jet \pt and mass by up to 30\% at large \npu, as compared to the Jet-by-jet CS correction. 
Across the full range of \npu considered, ICS exhibits an improvement beyond that of the CS correction alone by approximately 5-10\% in the resolution of both the jet 
\pt and the jet mass.

  \begin{figure}[!h]
  \centering
  \begin{subfigure}{0.47\textwidth}
    \includegraphics[width=\textwidth]{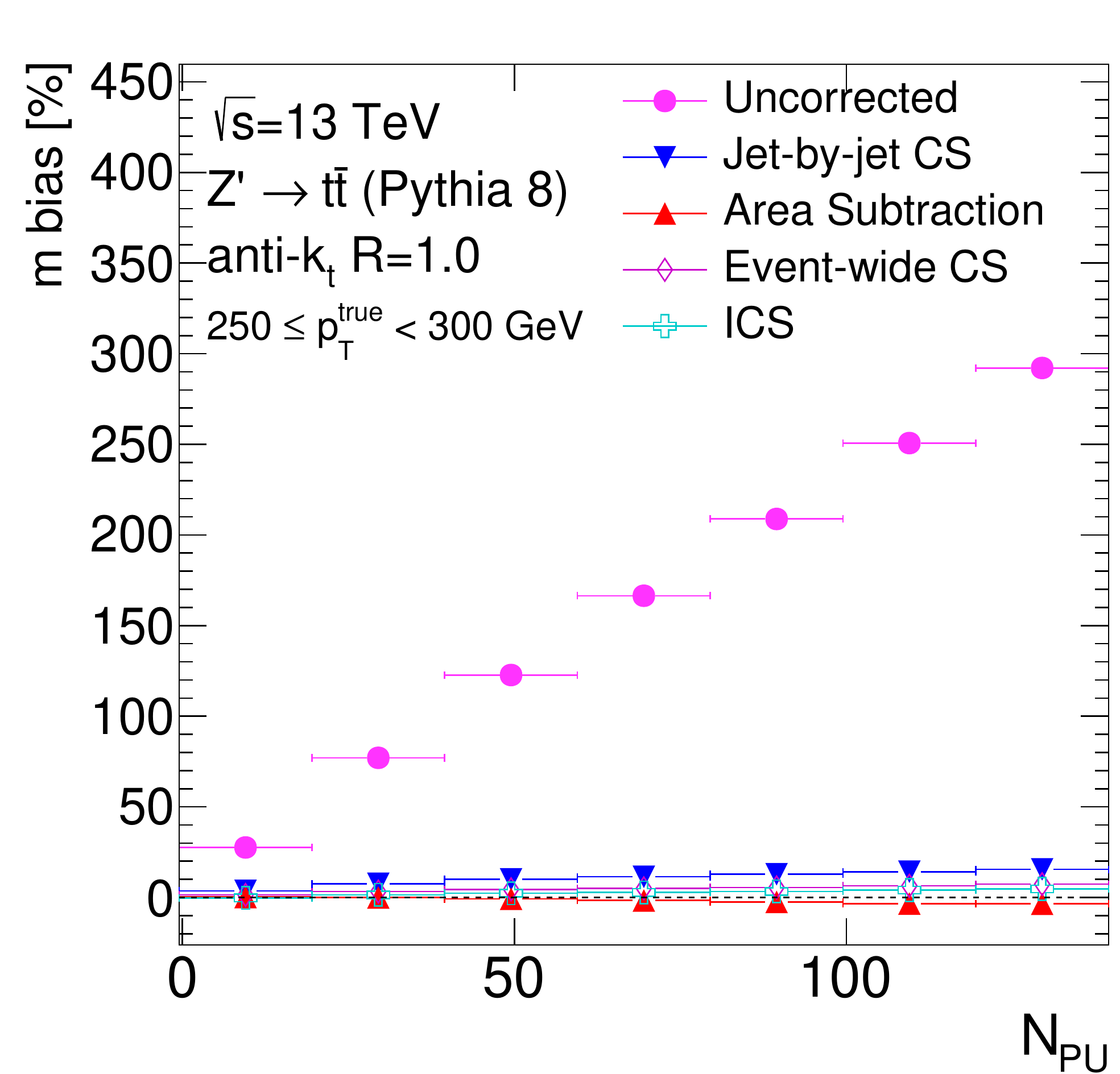}
    \caption{}
    \label{fig:Zprime:mass_bias}
  \end{subfigure}
  \begin{subfigure}{0.47\textwidth}
    \includegraphics[width=\textwidth]{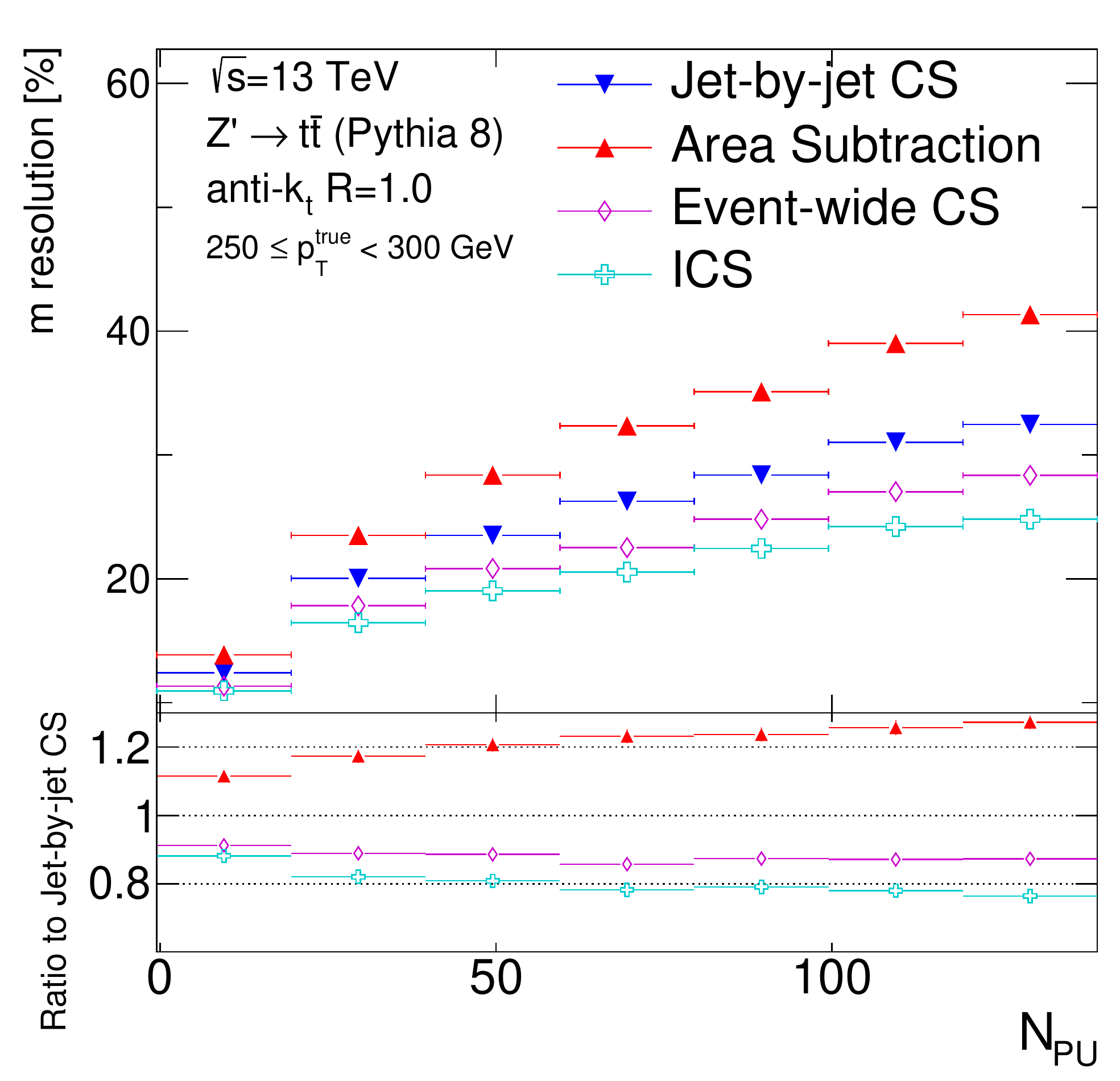}
    \caption{}
    \label{fig:Zprime:mass_resolution}
  \end{subfigure}
  \caption{Dependence of jet mass bias (left) and resolution (right) on \npu for four \pileup correction methods (Jet-by-jet CS, \areaSubtraction, Event-wide CS, 
and ICS).}
  \label{fig:Zprime:mass}
\end{figure}

\begin{figure}[!h]
  \centering
    \includegraphics[width=\textwidth]{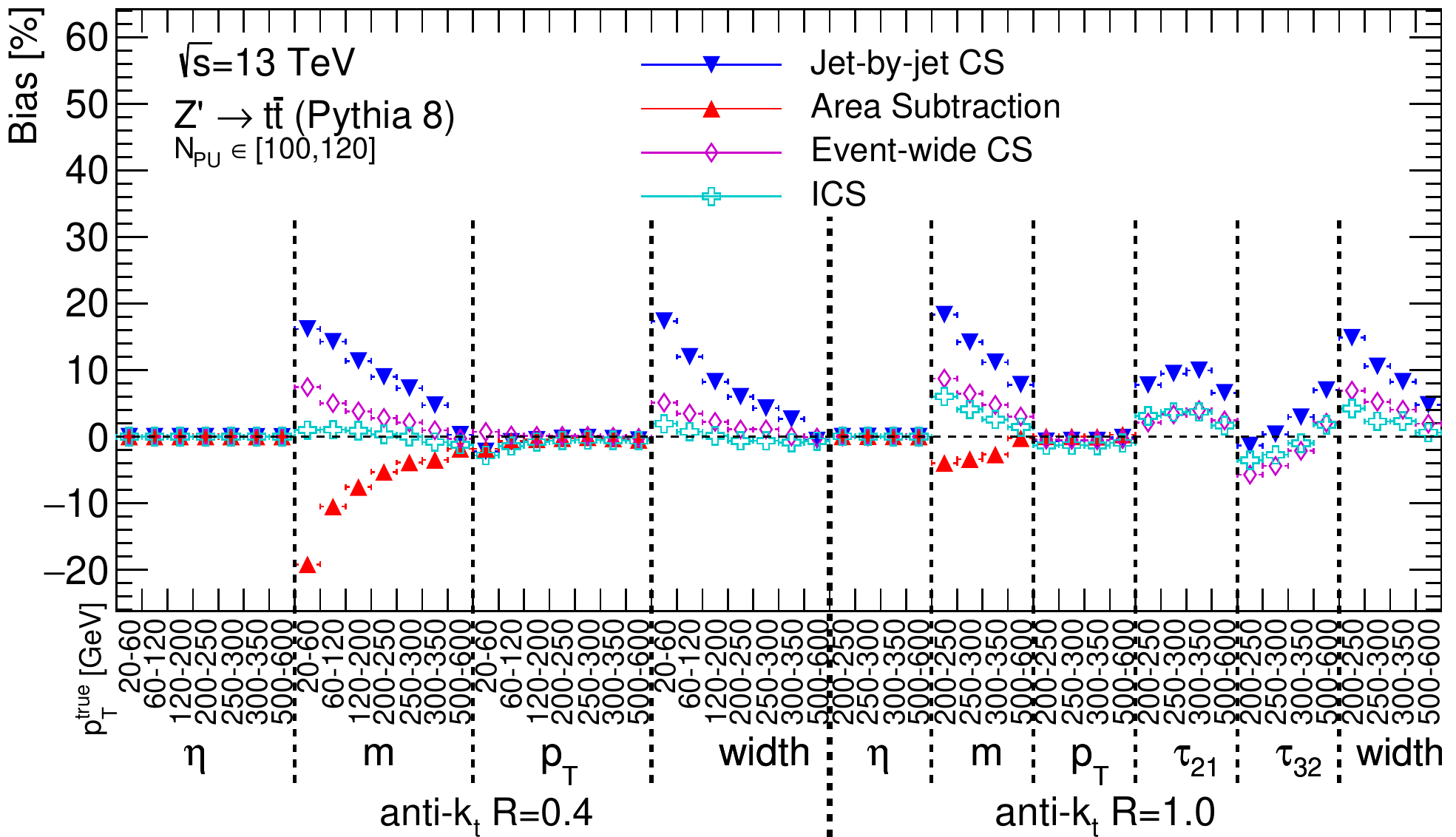}

  \caption{Performance of the \pileup subtraction evaluated in terms of the bias 
for different observables: jet $\eta$, mass, $\pt$, width for $R=0.4$
jets and jet $\eta$, mass, $\pt$, width, $\tau_{21}$, $\tau_{23}$ for
$R=1.0$ jets. Each bin on the $x$-axis represents a given range of true jet \pttrue\
defined by the bin-label. Bins on the $x$-axis are grouped to distinguish
observables and jet radii.
Four algorithms are compared: Jet-by-jet CS, \areaSubtraction, Event-wide CS, and ICS algorithm.
}  
  \label{fig:bias}
\end{figure}

The ability of each algorithm to mitigate the impacts of \pileup depends on more than just the amount of \pileup considered. As mentioned earlier, the size of the jet 
and the kinematic range (e.g. high or low true \pttrue) can affect the performance significantly. Moreover, the impact of \pileup on certain observables can be larger, 
thus reducing the effectiveness of certain approaches. \Figsref{bias}{resolution} summarize the results of a comprehensive study of the performance 
of each of the four algorithms under consideration. Results are reported in a narrow range of high \pileup, $\npu = 100\text{--}120$, for jets from the 
$Z' \rightarrow t\bar{t}$ process, for the following:

\begin{itemize}
  \item \textbf{jet radius:} $R=0.4, 1.0$
  \item \textbf{true jet \pt:} 
  \begin{itemize}
    \item 7 bins of \pttrue in the range $\pttrue = 20\text{--}600$~GeV for $R=0.4$
    \item 4 bins of \pttrue in the range $\pttrue = 200\text{--}600$~GeV for $R=1.0$
  \end{itemize}
  \item \textbf{observable:} 
  \begin{itemize}
    \item $\eta$, mass, \pt, width for both $R=0.4, 1.0$
    \item $\tau_{21}$ and $\tau_{32}$ for only $R=1.0$
  \end{itemize}
\end{itemize}

\noindent The outcome is a set of 52 comparisons each for the bias and resolution after \pileup correction. 

\begin{figure}[!h]
  \centering
    \includegraphics[width=\textwidth]{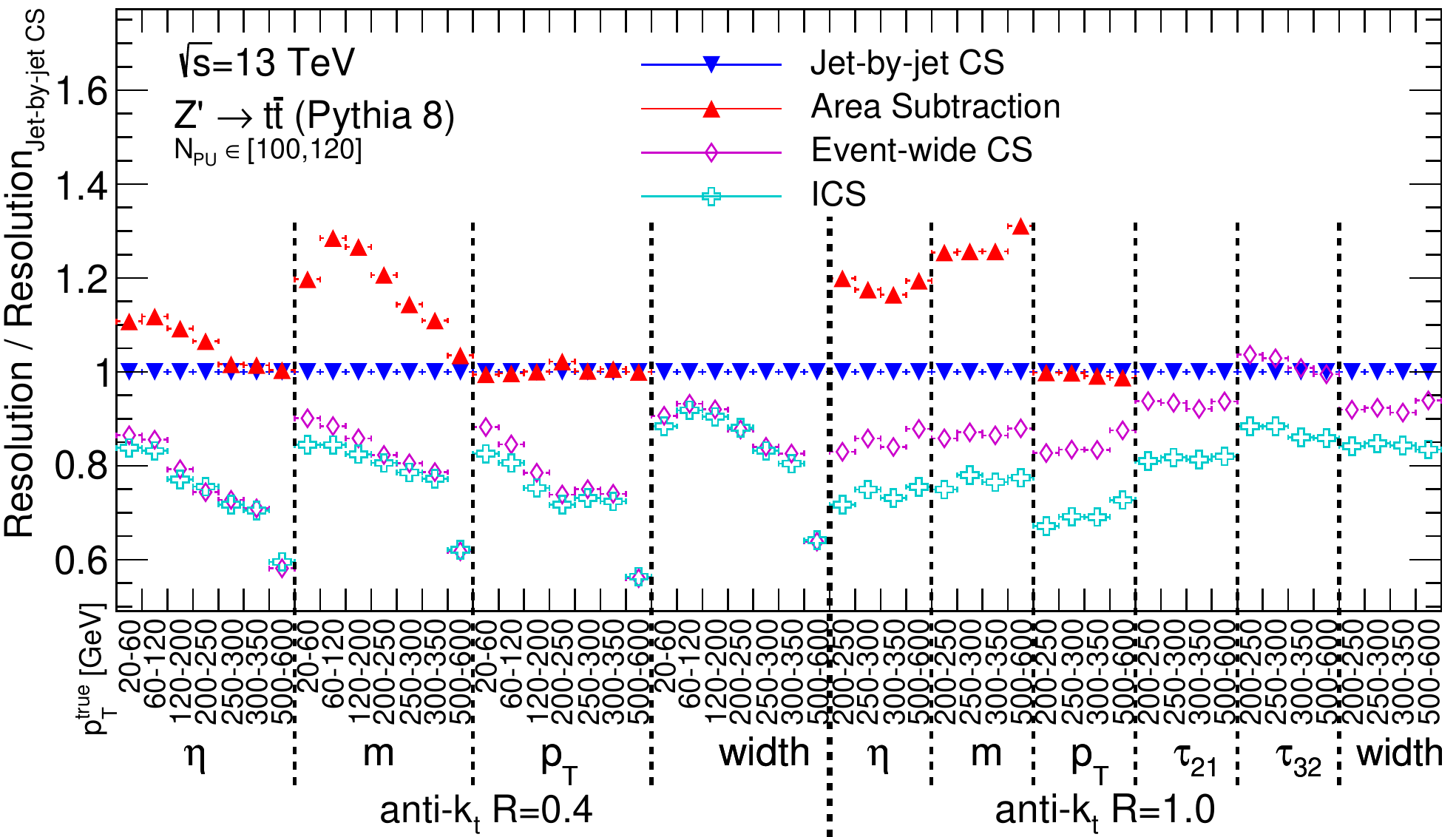}

  \caption{Performance of the \pileup subtraction evaluated in terms of the ratio of resolution with respect to the resolution from Jet-by-jet CS algorithm
for different observables: jet $\eta$, mass, \pt, width for $R=0.4$
jets and jet $\eta$, mass, \pt, width, $\tau_{21}$, $\tau_{23}$ for
$R=1.0$ jets. Each bin on the $x$-axis represents a given range of true jet \pttrue\
defined by the bin-label. Bins on the $x$-axis are grouped to distinguish
observables and jet radii. 
Four algorithms are compared: Jet-by-jet CS, \areaSubtraction, Event-wide CS, and ICS algorithm.
  }  
  \label{fig:resolution}
\end{figure}

The results of these comparisons provide a rich set of information from which a few conclusions may be reliably drawn. 
  The ability of algorithms to remove the bias is practically identical in the case of jet \pt and jet $\eta$. The ability to correct the bias is significantly 
improved for the Event-wide CS and ICS algorithms compared to Jet-by-jet CS algorithm in the case of jet mass, jet width, and $\tau_{21}$. For $\tau_{32}$, the 
performance of Jet-by-jet CS, Event-wide CS and ICS algorithms is very similar. The bias generally decreases with increasing jet \pttrue\ and tend to converge to zero 
for all the algorithms, both jet radii, and almost all the observables. Such a convergence in the behavior of algorithms is however not seen in the case of resolution, 
where significant differences among algorithms persist in the full kinematic range and in some cases even tend to get larger at higher \pttrue.

A clear improvement in the resolution is seen for Event-wide CS and ICS algorithms compared to both Jet-by-jet CS algorithm and \areaSubtraction. While the 
resolution from ICS and Event-wide CS is practically the same for $R=0.4$ jets, a significant difference is seen for $R=1.0$ jets where ICS algorithm outperforms the 
Event-wide CS. These observations hold for all observables studied and for all the jet \pttrue\ bins. This conclusion together with the conclusion on the bias implies 
that ICS algorithm provides the largest and most consistent improvements in the performance out the four algorithms tested.

\subsection{Performance for missing transverse energy}
\label{sec:missinget}
In addition to evaluating the performance of ICS and related background subtraction algorithms in terms of their impacts on jets, we also studied the extent to which improvements might be gained in the measurement of missing transverse energy, \etmissVec. The \etmissVec in the event is defined as a 2-vector calculated as the negative vector sum of the \pt of all physics objects in the event. For this study, \etmissVec is calculated using the vector sum of all stable particles, while jets are not used in the calculation. The results are reported in terms of just one component of the 2-vector, \etmissx. 

\begin{figure}[!h]
  \centering
    \includegraphics[width=0.6\textwidth]{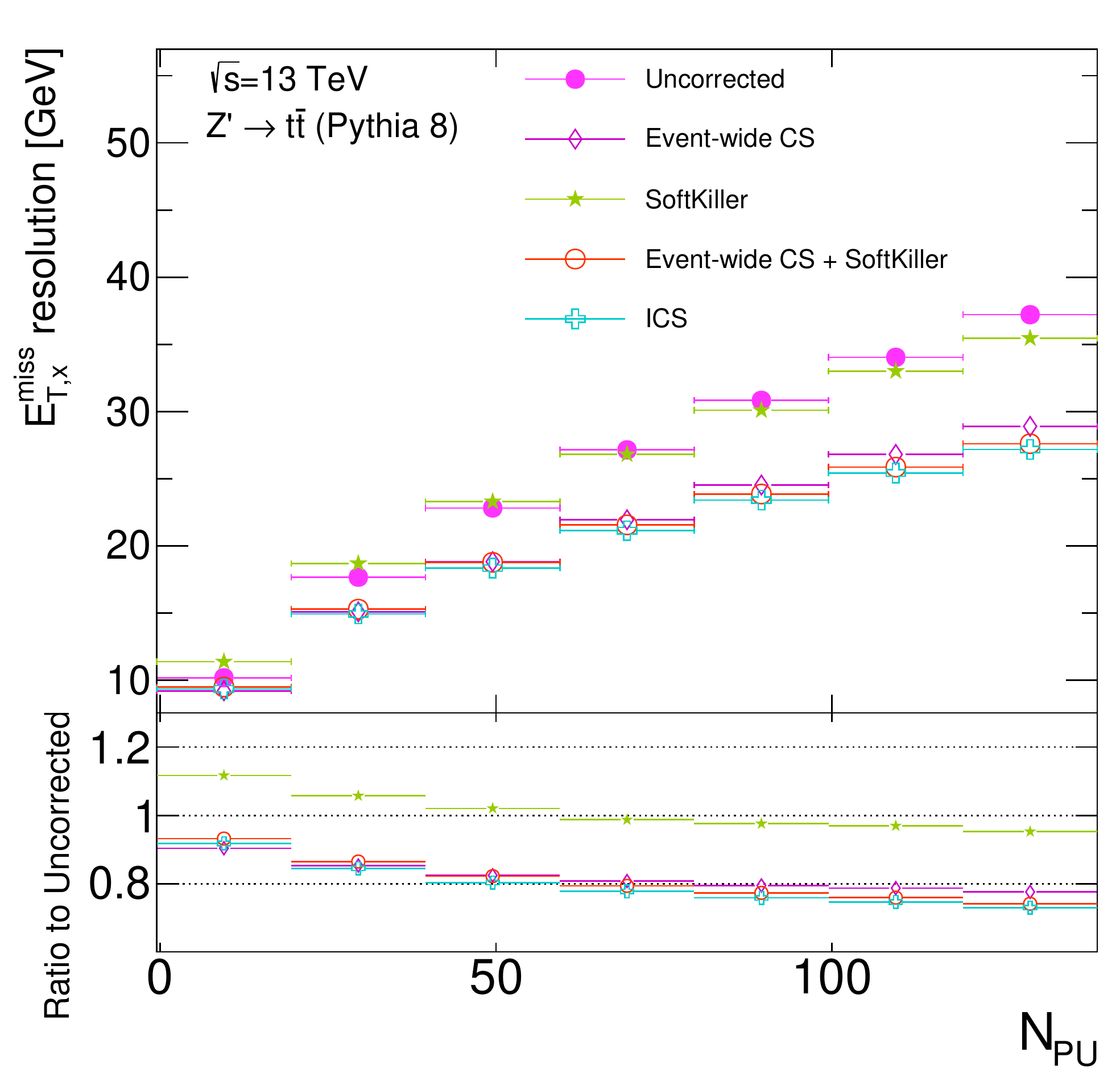}
  \caption{Resolution of the $x$-component of the \etmissVec.}
  \label{fig:MET_resolution}
\end{figure}

\Figref{MET_resolution} shows as a function of \npu a measure of the resolution of the \etmissx, calculated as the RMS of the difference between the reconstructed and 
true \etmissx in the event. The uncorrected resolution is reported as well as the results of applying four different subtraction algorithms to the entire event: 
SoftKiller (using grid-size parameter of $0.6$), Event-wide CS and ICS with the same configurations as for \aktfour jets described in \secref{performance_setup}, and 
combination of Event-wide CS followed by SoftKiller (tested by the ATLAS Collaboration \cite{ATLAS:2017pfq}). The optimal grid-size parameter for SoftKiller depends on 
the $\npu$. For example, grid-size parameter of $0.5$ provides by ${\sim}5\%$ better resolution for $\npu<60$, but by ${\sim}5\%$ worse resolution for $\npu>100$ 
compared to the configuration with the grid-size of $0.6$. The grid-size parameter of $0.7$ provides by ${\sim}3\%$ better resolution for $\npu > 100$, but by 
${\sim}10\%$ worse resolution for $\npu <60$ compared to the configuration with the grid-size of $0.6$. For this reason we show SoftKiller with the grid-size parameter 
of 0.6 which is a good compromise. We found that the best resolution for the CS and SoftKiller combination can be achieved by using CS parameters $\DeltaRmax=0.2$ and 
$\alpha=1$, and SoftKiller grid-size parameter of 0.5.

For all the tested methods, the resolution of the \etmissx worsens with increasing \pileup, as expected. However, the three methods Event-wide CS, ICS, and Event-wide CS followed by SoftKiller reduce this degradation by more than $20\%$ at large \npu, while SoftKiller alone does not perform so well. This along with results presented in the previous section demonstrate the stability and good performance of the ICS method.

\subsection{Performance using the framework from the 2014 Pileup Workshop}
\label{sec:performance_workshop}
We compare the performance of the new method with other methods using the common open-source software framework~\cite{PileupWorkshopFramework} defined at the ``Pileup Workshop'' held in May 2014 at CERN~\cite{PileupWorkshopIndico}. The code used to obtain the results in this section is located in the folder \texttt{comparisons/ICS} of this framework, version 1.1.0. 

  The samples used for the comparison presented here are available from \bibref{PileupWorkshopSamples}. Four hard-scatter physics processes overlaid with a fixed number of \pileup events are used. The four hard-scatter samples are dijet events with at least one reconstructed \aktfour jet satisfying $\pt\geq$ 20, 50, 100 or 500\gev. $B$-hadrons are kept stable and UE is not simulated. The hard-scatter and \pileup samples are simulated using Pythia 8.185 with tune 4C using proton-proton collisions at $\sqrt{s}=14\tev$. Four \pileup conditions are used: $\npu=$ 30, 60, 100, and 140.  The performance is evaluated for \aktfour jets.

The mass of all particles is set to zero, preserving \pt, rapidity, and azimuth. Only particles with $|y|<4$ are used. The corrections are done on events without any detector simulation but assuming idealised tracker, where the information about 
the origin of charged particles (\pileup or hard-scatter event) is known. In that way, the charged \pileup particles can be directly removed and/or the information 
about charged particles can be further used to correct the neutral particles. We compare the CS-based methods with \areaSubtraction, SoftKiller, and PUPPI, for which we use the same configurations as for the comparison in \bibref{Soyez:2018opl} except slightly different $\rho$ estimation for the \areaSubtraction as described below. 
Summary of the used configurations and usage of the information about charged particles is the following:
  \begin{itemize}
  \item \textbf{\areaSubtraction:} All charged \pileup particles are discarded. The $\rho$ is estimated using only neutral particles with a grid-size parameter of 0.6 and rapidity rescaling. The protection against negative masses after subtraction is enabled. To determine the jet area, the ghosts are placed up to the edge of 
the particle rapidity acceptance, $|y|<4$, with a ghost area of 0.01.

  \item \textbf{SoftKiller:} All charged \pileup particles are discarded at the beginning. SoftKiller with grid-size parameter of 0.5 is applied on the neutral 
particles in the event. The used grid-size parameter is the same as in the \texttt{comparisons/review} in \bibref{PileupWorkshopFramework} version 1.0.0 which was used 
for the publication \cite{Soyez:2018opl}.\footnote{We have investigated the performance of other grid-size parameters and confirmed that the value of 0.5 represents the 
optimal choice.}

  \item \textbf{PUPPI:} The implementation is taken from the \texttt{comparisons/review} in \bibref{PileupWorkshopFramework} version 1.0.0, which is the original implementation provided by the PUPPI authors in the context of the 2014 Pileup Workshop.

  \item \textbf{Jet-by-jet CS:} All charged \pileup particles are discarded. Same $\rho$ estimation as for the \areaSubtraction is used. 
Only neutral particles are corrected using CS with parameters: $\DeltaRmax=\infty$, $\alpha=0$, and $\Aghost=0.01$.

  \item \textbf{Event-wide CS:} All charged \pileup particles are discarded. Same $\rho$ estimation as for the \areaSubtraction is used. Only the neutral particles are corrected using the Event-wide CS with parameters: $\DeltaRmax=0.25$, $\alpha=1$, and $\Aghost=0.005$.

  \item \textbf{CS+SoftKiller:} All charged \pileup particles are discarded. First Event-wide CS is applied as described in the previous point (with the same CS 
parameters). Then the corrected neutral particles are further corrected using SoftKiller (grid-size parameter of 0.6).\footnote{We have investigated the performance for
other CS parameters and SoftKiller grid-size parameter and identified this to be the optimal configuration.}

  \item \textbf{ICS:} All charged \pileup particles are discarded. Same $\rho$ estimation as for the \areaSubtraction. Only the neutral particles are corrected using the ICS method with two iterations and without ghost removal. Parameters: $\alpha=1$ and $\Aghost=0.005$ for both iterations, $\DeltaRmax=0.2$ and $\DeltaRmax=0.15$ for the first and the second iteration, respectively.

  \end{itemize}

\texttt{FastJet} v3.3.2 is used for jet clustering, background estimation and \areaSubtraction. \texttt{FastJet Contrib} v1.038 is used for CS-based methods and 
SoftKiller.
  To compare the performance of the methods, the average bias and resolution for the jet \pt and jet mass are evaluated. From each event, the two true hardest jets are 
selected with $\pt>20\gev$ and $|y|<2.5$. Each corrected jet is matched to the closest true jet requiring $\sqrt{\Delta y^2+\Delta\phi^2}<0.3$ (the matching efficiency 
is above $99.5\%$).
  To follow recommendations from the workshop, the average jet \pt bias, $\langle\Delta\pt\rangle$, and \pt resolution, $\sigma_{\Delta\pt}$, is evaluated as the average 
and RMS from the \pt difference between true and matched corrected jet \pt, respectively. Similary, the mass bias, $\langle\Delta m\rangle$, and resolution, 
$\sigma_{\Delta m}$, are evaluated.

\begin{figure}[!h]
  \centering            
  \includegraphics[width=0.79\textwidth,origin=c]{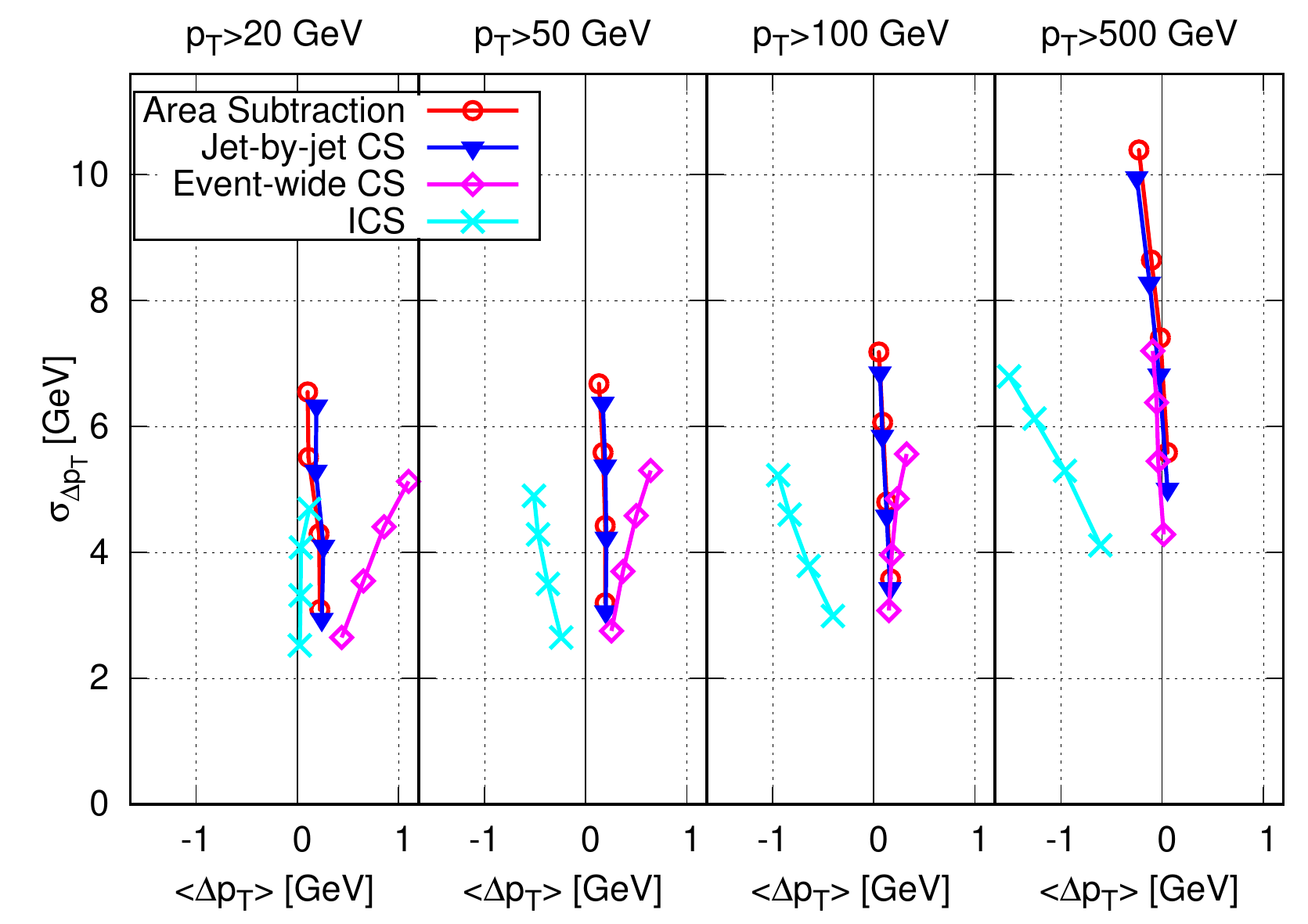}
  \caption{Comparison of the jet \pt resolution as a function of jet \pt bias for the \areaSubtraction, jet-by-jet CS, Event-wide CS and ICS methods. Dijet events are used with different jet \pt cut in each panel. Each curve corresponds to a different method and the 4 points on each curve correspond to $\npu=$ 30, 60, 100 and 140 from bottom to top.}  
  \label{fig:workshop_pt}
\end{figure}

The performance of the \areaSubtraction, Jet-by-jet CS, Event-wide CS and ICS methods is shown in \figsref{workshop_pt}{workshop_mass}. Qualitatively, the differences 
between the individual methods are the same as presented in \secref{jet} where, however, a different signal sample and detector simulation are used. Both, Event-wide CS 
and ICS, improve the resolution significantly with respect to the \areaSubtraction and Jet-by-jet CS. The ICS method has slightly better resolution with smaller biases 
compared to the Event-wide CS.

\begin{figure}[!h]
  \centering
  \includegraphics[width=0.79\textwidth,origin=c]{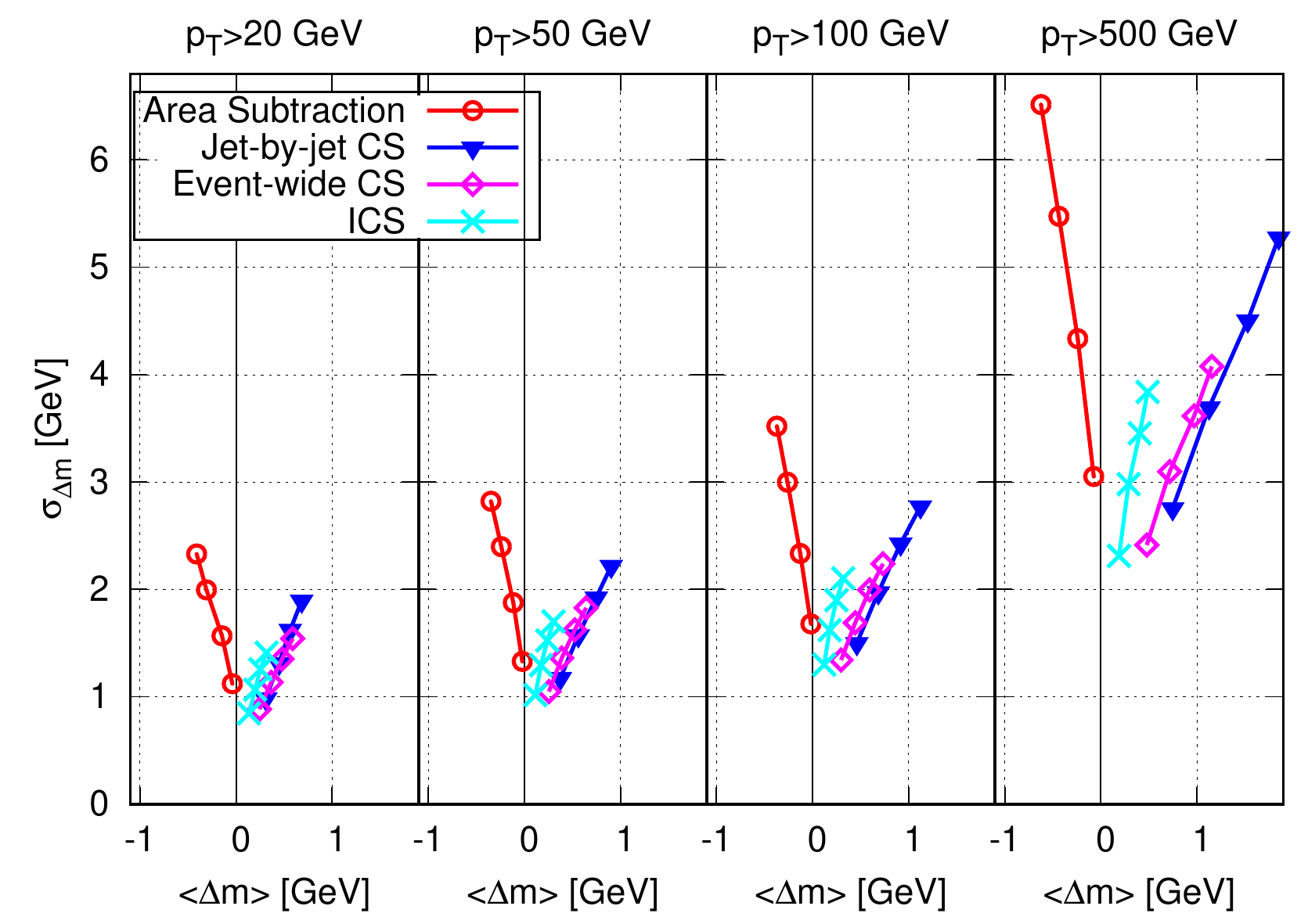}
  \caption{Comparison of the jet mass resolution as a function of jet mass bias for the \areaSubtraction, jet-by-jet CS, Event-wide CS and ICS methods. Dijet events are used with different jet \pt cut in each panel. Each curve corresponds to a different method and the 4 points on each curve correspond to $\npu=$ 30, 60, 100 and 140 from bottom to top.}  
  \label{fig:workshop_mass}
\end{figure}

\begin{figure}[!h]
  \centering            
  \includegraphics[width=0.79\textwidth,origin=c]{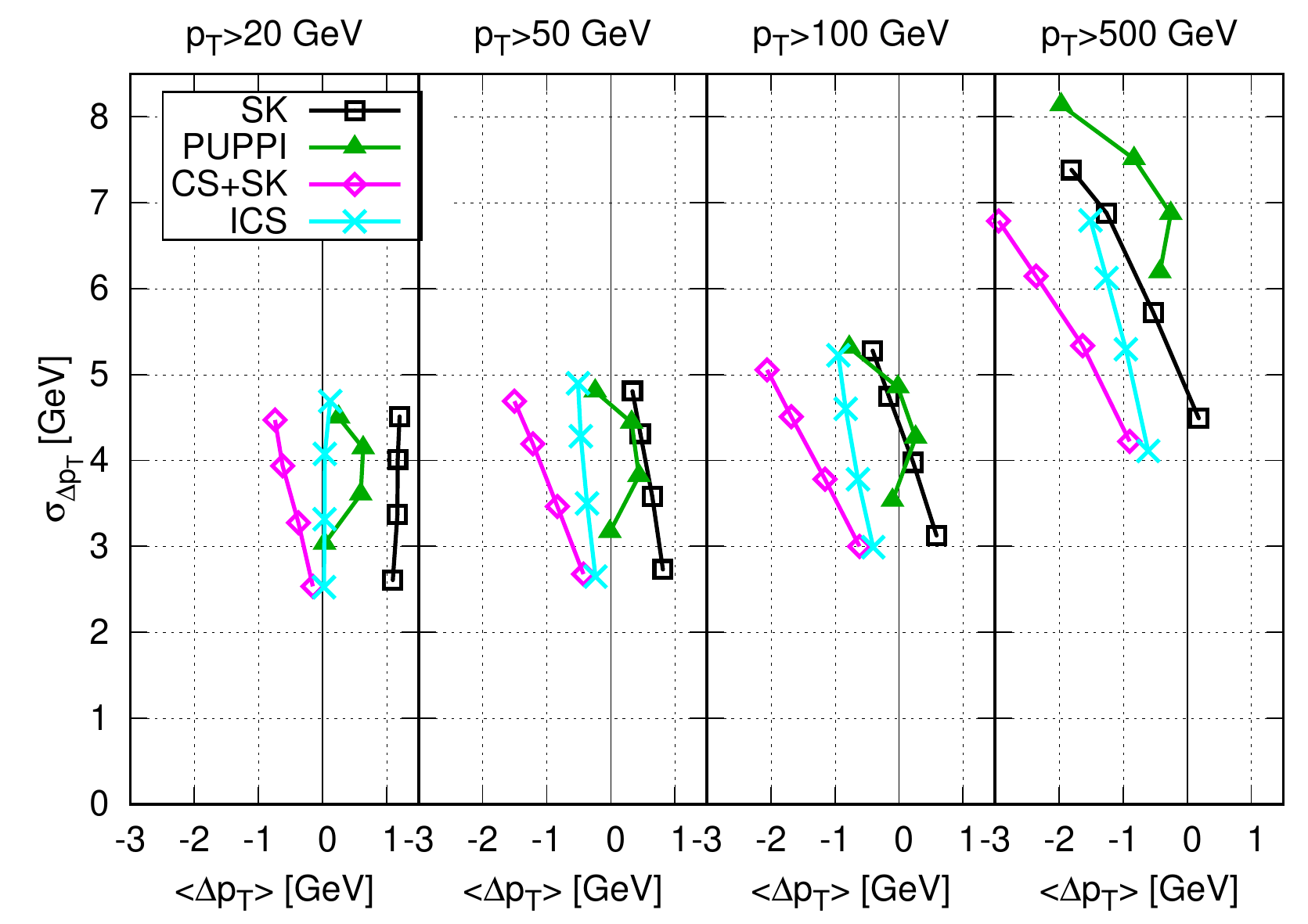}
  \caption{Comparison of the jet \pt resolution as a function of jet \pt bias for four methods: SoftKiller, PUPPI, CS+SoftKiller and ICS. Dijet events are used with different jet \pt cut in each panel. Each curve corresponds to a different method and the 4 points on each curve correspond to $\npu=$ 30, 60, 100 and 140 from bottom to top.}  
  \label{fig:workshop_pt2}
\end{figure}

The performance of the PUPPI, SoftKiller, CS+SoftKiller, and ICS methods is shown in \figsref{workshop_pt2}{workshop_mass2}. All the compared methods lead to varying 
levels of bias for some observables or kinematic selections. For the majority of the \npu and kinematic selections, the best performance is achieved by the ICS 
method. For all the \npu and kinematic selections, PUPPI provides slightly worse resolution of \pt\ compared to ICS which is worsening with increasing \pt. On the contrary, for 
high \npu, the performance of PUPPI is systematically better compared to ICS in terms of the reconstruction of mass.
  In the low \npu environment the performance of ICS and CS+SoftKiller is almost identical, but for a higher \npu environement, CS+SoftKiller gives larger negative bias. 
On the contrary, SoftKiller alone gives systematically positive bias. This oversubtraction or undersubtraction of SoftKiller might be avoided by further optimizing the 
SoftKiller parameters to avoid cutting out a part of the signal or to allow for cutting out more of the background. The SoftKiller performance may also be improved by 
applying the \textit{protected zeroing} which removes all neutral particles below certain \pt\ threshold if a given particle is not located close to a charged particle 
from a hard-scattering process as discussed in \bibref{Soyez:2018opl}.
  We should emphasize that in this study, the only method which uses the information about charged particles in the subtraction is PUPPI. In general, using 
the information from charged particles is expected to improve the performance of the subtraction. Systematic study of the methods which use the information from charged 
particles within the context of constituent-subtraction-based algorithms goes beyond the scope of this paper and is planned for a separate study.

\begin{figure}[!h]
  \centering
  \includegraphics[width=0.79\textwidth,origin=c]{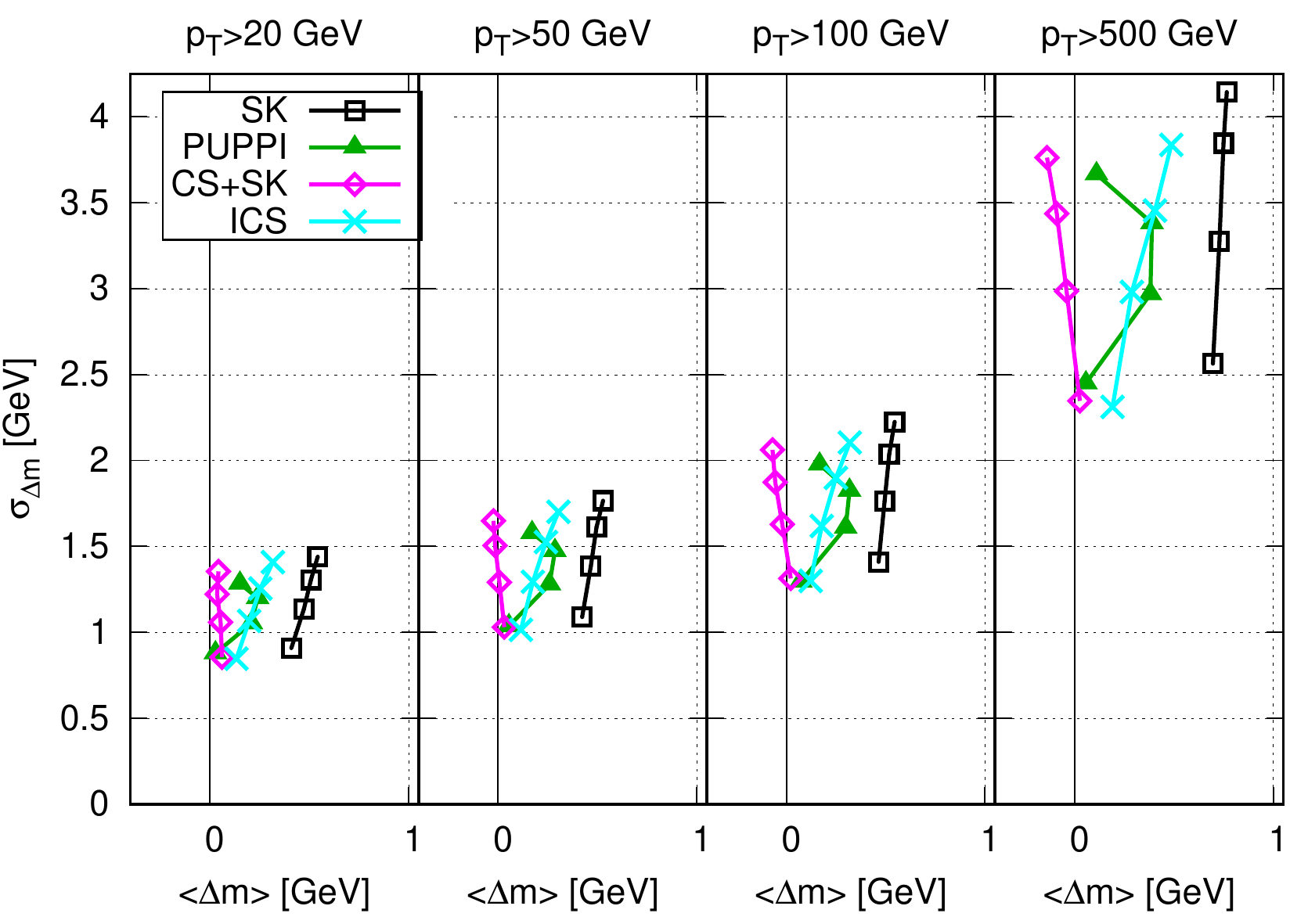}
  \caption{Comparison of the jet mass resolution as a function of jet mass bias for four methods: SoftKiller, PUPPI, CS+SoftKiller and ICS. Dijet events are used with different jet \pt cut in each panel. Each curve corresponds to a different method and the 4 points on each curve correspond to $\npu=$ 30, 60, 100 and 140 from bottom to top.}  
  \label{fig:workshop_mass2}
\end{figure}

\section{Conclusions}
\label{sec:conclusions}
We presented a new background mitigation method for jet kinematics, jet substructure observables, and event observables (such as missing transverse energy), called Iterative Constituent Subtraction. This method is applicable to both \pileup effects in proton-proton collisions and underlying event contributions in heavy-ion collisions. Iterative Constituent Subtraction extends the Constituent Subtraction method from \bibref{Berta:2014eza} to an iterative event-wide subtraction algorithm with improved features and performance. The new algorithm has been tested using hadronic jets from $Z' \rightarrow t\bar{t}$, as well as light quark and gluon dijet processes, and it was compared to various other algorithms including Jet-by-jet CS, Event-wide CS, \areaSubtraction, SoftKiller, and PUPPI. Iterative Constituent Subtraction has been shown to significantly improve the performance of  \pileup subtraction in proton-proton collisions in terms of bias and resolution of the jet kinematics and substructure observables compared to the \areaSubtraction, Jet-by-jet CS, and Event-wide CS. The improvement was also observed with respect to other methods for a large number of \pileup and kinematic configurations. The new method has potential to improve the background mitigation at both proton-proton  and heavy-ion colliders such as the LHC and RHIC.

\appendix
\section{General discussion on the CS parameters}
\label{app:CS_parameters}
As discussed in \secref{algo}, the CS procedure has three free parameters to control the subtraction: \DeltaRmax, $\alpha$, and \Aghost. These parameters have weak 
impact in the Jet-by-jet CS, and the recommended values in that configuration are $\DeltaRmax=\infty$, $\alpha=0$, and $\Aghost\leq0.01$. On the contrary, the 
Event-wide CS is much more sensitive to the \DeltaRmax parameter and also the $\alpha$ parameter can have non-negligible effect. This section serves as a general 
guidance how to set the three parameters for the Event-wide CS. However, the conclusions found in our simulation may not be perfect for a specific detector environment. Therefore the experiments are encouraged to optimize the parameters in their own environment. The recommended starting point of tests for Event-wide CS is: $\DeltaRmax=0.25$, $\alpha=1$, $\Aghost = 0.0025$ for \aktfour jets and $\DeltaRmax=0.7$, $\alpha=1$, $\Aghost = 0.0025$ for \aktten jets. We provide justification for these recommendations and a basic analysis of the sensitivity to the choice of parameters in the following sub-sections. The performance studies presented here use the setup described in \secref{performance_setup}.

\subsection{Maximal distance between ghost-particle pairs, \DeltaRmax}
\label{app:DeltaRmax}

The parameter \DeltaRmax controls the maximal allowed distance between ghost-particle pairs. Only ghost-particle pairs which have $\DeltaR<\DeltaRmax$ are combined in 
the algorithm. By setting a finite \DeltaRmax, one can avoid combining ghost-particle pairs which are too far from each other. However, in this case, certain amount of 
estimated \pileup can remain in the event, while by setting $\DeltaRmax=\infty$ it is ensured that all the estimated \pileup represented by ghosts is subtracted from 
particles, although not necessarily at a naturally small distance between ghosts and particles.

\begin{figure}[!h]
  \centering
  \includegraphics[width=0.5\textwidth]{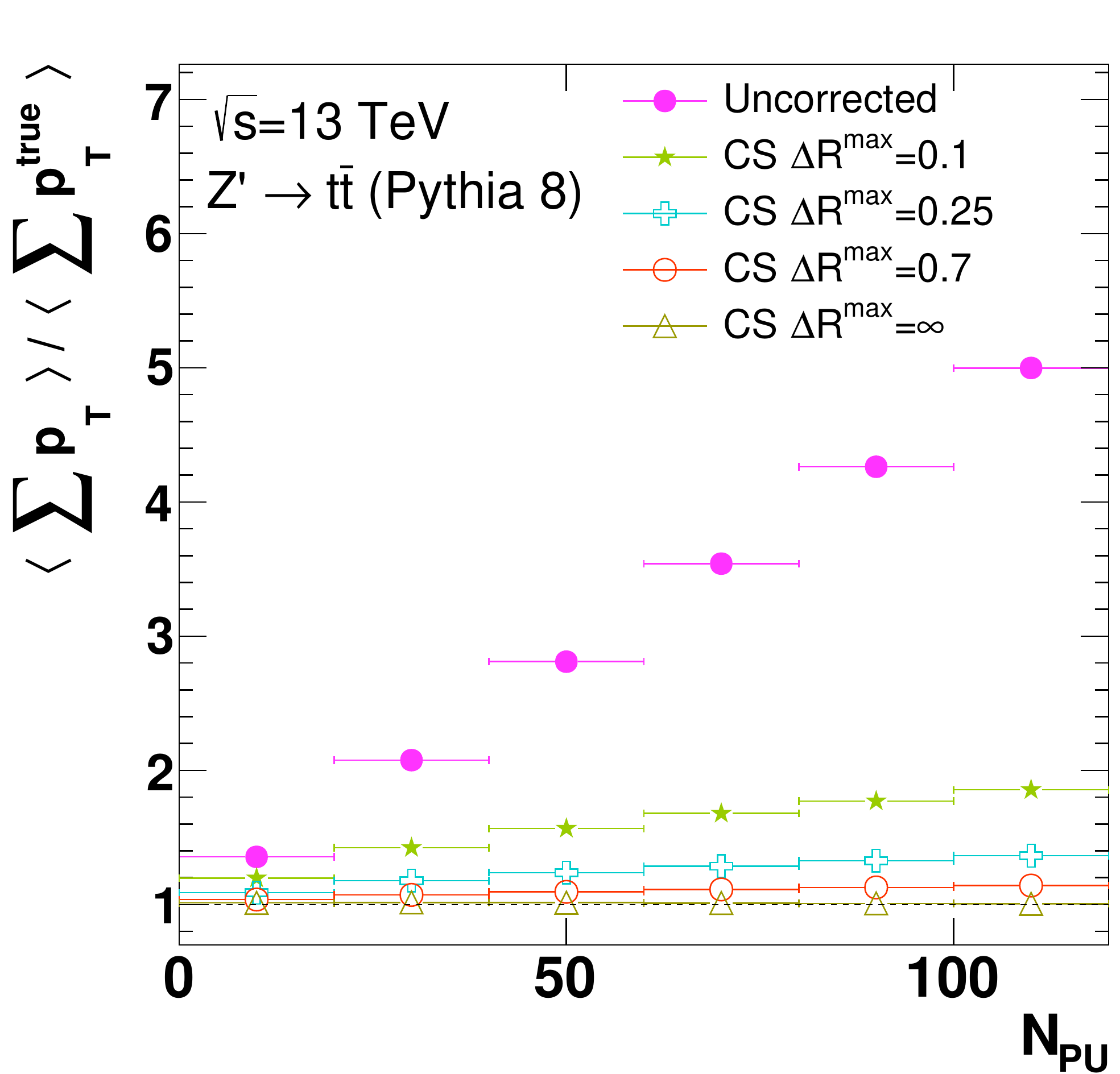}
  \caption{Ratio between the \ptSum for various \DeltaRmax parameters used in the Event-wide CS and \ptSumTrue as a function of \npu. }
  \label{fig:maxDeltaR_ptSum}
\end{figure}

We investigated the amount of remaining \pileup for various \DeltaRmax values by evaluating the scalar \pt sum from all particles in the event. The average scalar \pt 
sum for true events (events without \pileup) for the used physics process is $\ptSumTrue\approx 1 \tev$. The average scalar \pt sum for the same events with \pileup 
after correction, \ptSum, varies depending on the \DeltaRmax parameter as shown in \figref{maxDeltaR_ptSum}. For $\DeltaRmax=\infty$, the $\ptSum\approx \ptSumTrue$ for 
the whole \npu range, which means that on average, the expected \pileup deposition is entirely subtracted. This is expected, but this evaluation represents an important 
cross-check to see that the used background estimation is not biased. For finite \DeltaRmax, certain amount of \pileup remains in the events. For example, there is 
still a remaining average background \pt of ${\sim}80\mev$ per unit area per one \pileup event when using $\DeltaRmax=0.25$.

\begin{figure}[!h]
  \centering
  \includegraphics[width=\textwidth]{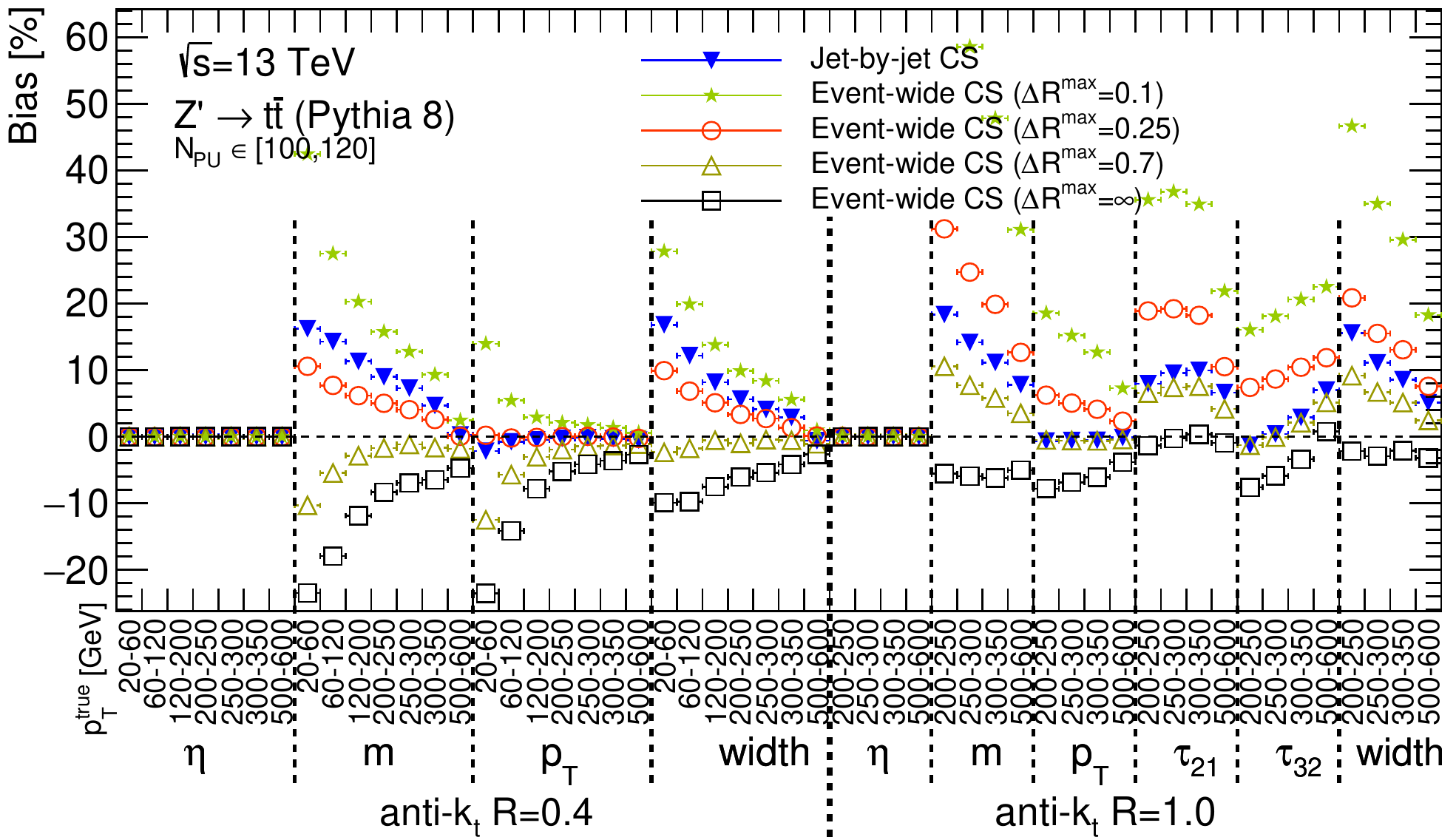}
  \caption{Dependence of bias on the \DeltaRmax parameter for the Event-wide CS. The other CS parameters are $\alpha=0$ and $\Aghost=0.0025$.}
  \label{fig:maxDeltaR_bias}
\end{figure}

The justification for using small values of \DeltaRmax comes from the actual performance on jets, see \figsref{maxDeltaR_bias}{maxDeltaR_resolution}. When using 
$\DeltaRmax=\infty$, the jets are largely overcorrected. The reason for this are the fluctuations of \pileup in the \yphi space. In case of no fluctuations, the ghosts 
subtract the \pt added by \pileup rather accurately. However, the presence of fluctuations can cause that ghosts are more often matched to a hard-scatter particle, 
which may be distant, which leads to the overcorrection.

The optimal \DeltaRmax value may depend on the jet definition and the detector granularity. We found that $\DeltaRmax=0.25$ leads to optimal performance 
for both \aktfour and \aktten keeping the biases low and maximizing the resolution for most observables. Using $\DeltaRmax=0.7$ can lead to lower biases for \aktten 
jets in expense of worse resolution. In any case, it is recommended to apply an additional correction to remove the remaining \pileup which can be addressed by the \ics 
method described in \secref{ics}.  Other possibility is to use a different method. For example, SoftKiller (grid parameter of 0.6) applied after the Event-wide CS with $\DeltaRmax=0.25$ was found to be one of the best methods in studies from the ATLAS Collaboration \cite{ATLAS:2017pfq}.

\begin{figure}[!h]
  \centering
  \includegraphics[width=\textwidth]{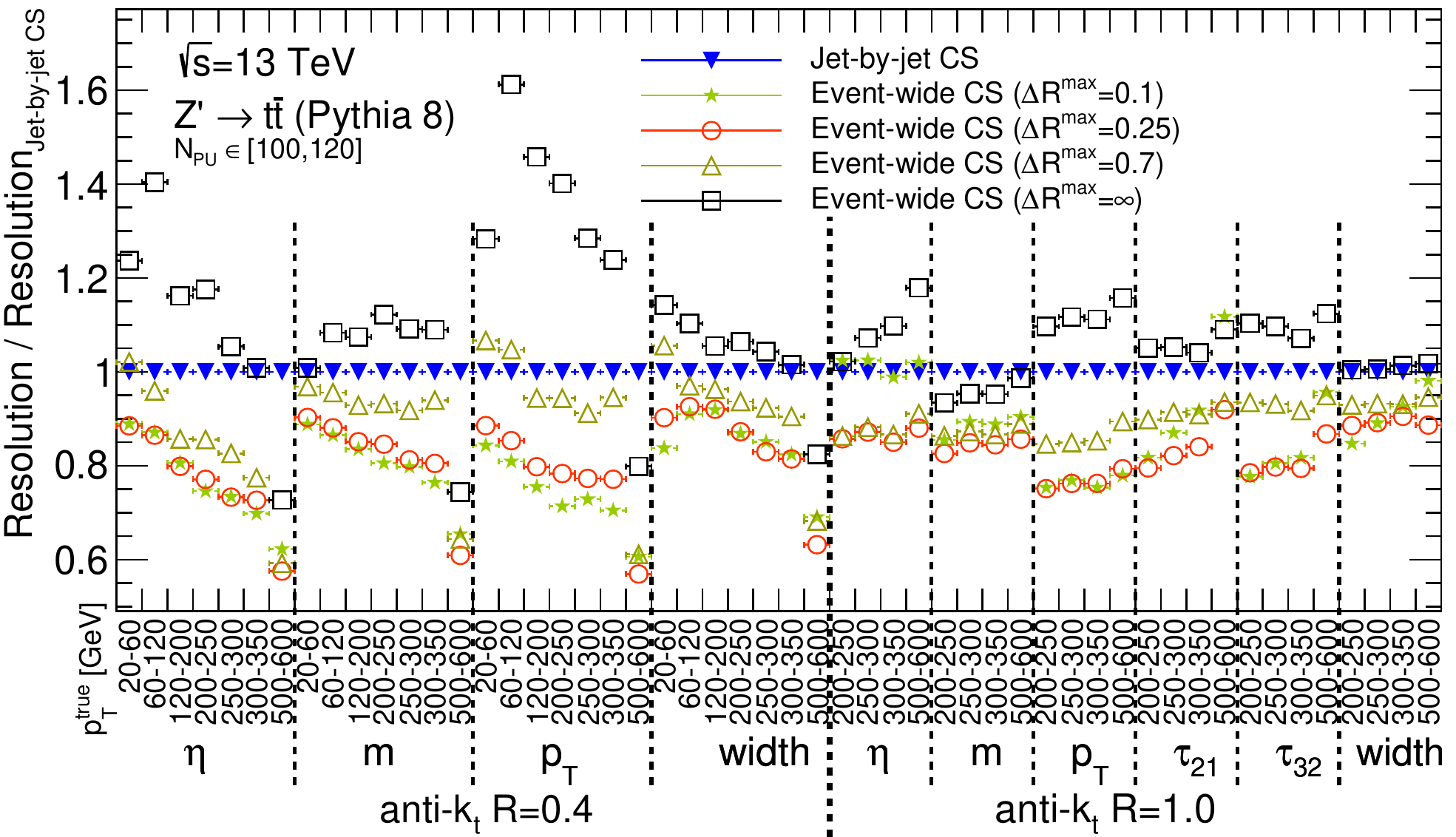}
  \caption{Dependence of resolution on the \DeltaRmax parameter for the Event-wide CS. The other CS parameters are $\alpha=0$ and $\Aghost=0.0025$.}
  \label{fig:maxDeltaR_resolution}
\end{figure}

\subsection{$\alpha$ parameter}

By using $\alpha>0$, one can prioritize ghost-particle pairs with lower particle \pt during the subtraction procedure. This may have positive effect on the performance 
since it is expected that the \pileup particles have lower \pt than the hard-scatter particles. The optimal value of the $\alpha$ parameter depends on many factors: 
granularity of the detector, possible \pt cuts applied to all particles, jet definition, \DeltaRmax parameter and also on the observable in question.

\begin{figure}[!h]
  \centering
  \includegraphics[width=\textwidth]{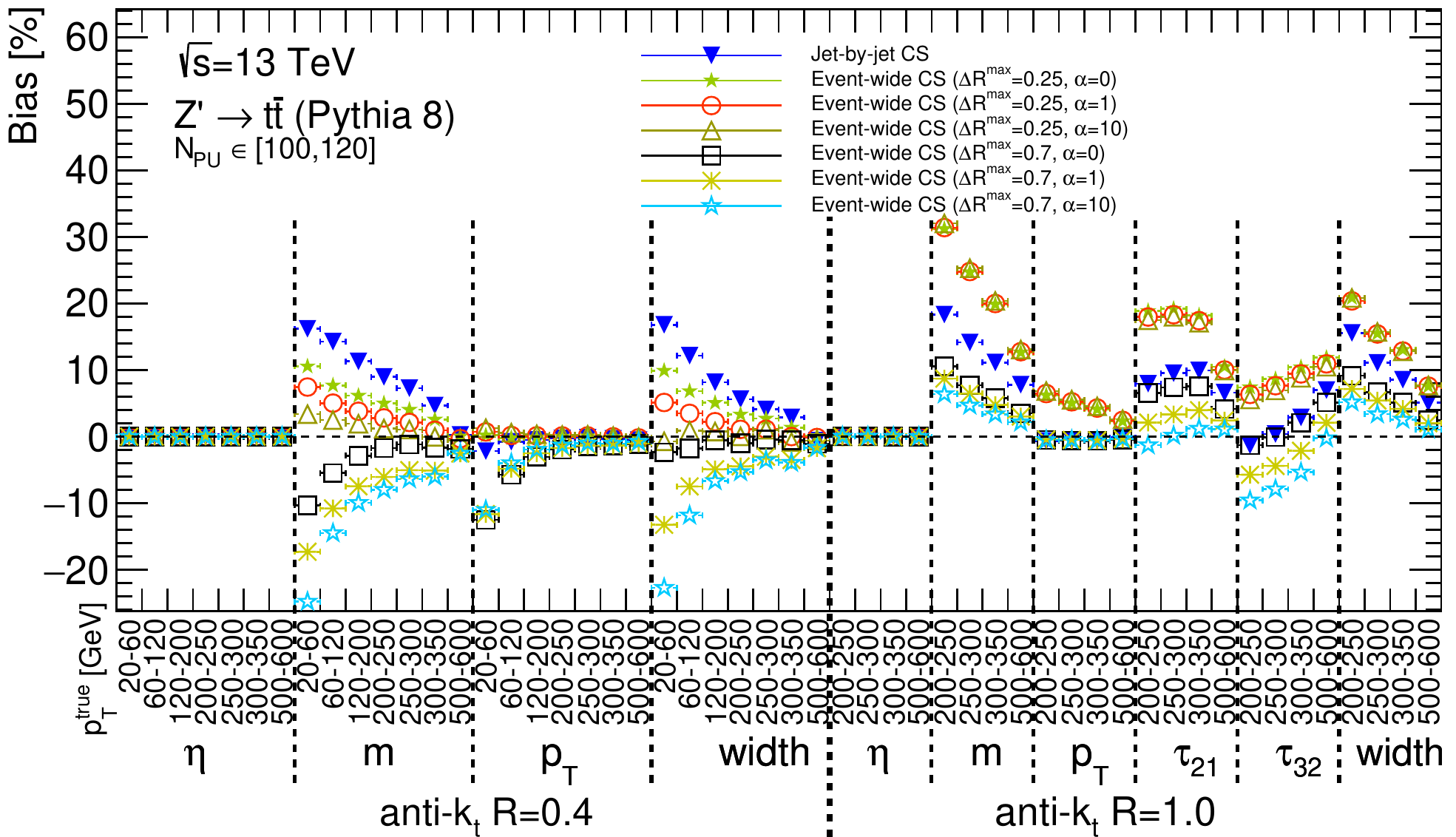}
  \caption{Dependence of bias on the $\alpha$ parameter for the Event-wide CS.}
  \label{fig:alpha_bias}
\end{figure}

The dependence of the Event-wide CS performance on $\alpha$ is shown in \figsref{alpha_bias}{alpha_resolution}. In general, the smaller the \DeltaRmax, the smaller 
is the effect of the $\alpha$ parameter. This is expected since with smaller \DeltaRmax, each ghost has smaller freedom to move to match with a particle during the CS 
procedure. The choice of $\alpha$ parameter has almost no effect for the \aktten jets when using small values of $\DeltaRmax$ (e.g. $\DeltaRmax=0.25$). In 
this case, each ghost is active over an area which is much smaller than the area of the jet, that is locally, and the choice of $\alpha$ has no real impact.

We found that the choice $\alpha=1$ for \aktfour and $\DeltaRmax=0.25$ keeps the biases low while the resolution is maximized. 
For \aktten jets, $\DeltaRmax=0.7$ and $\alpha=1$ is preferable.

\begin{figure}[!h]
  \centering
  \includegraphics[width=\textwidth]{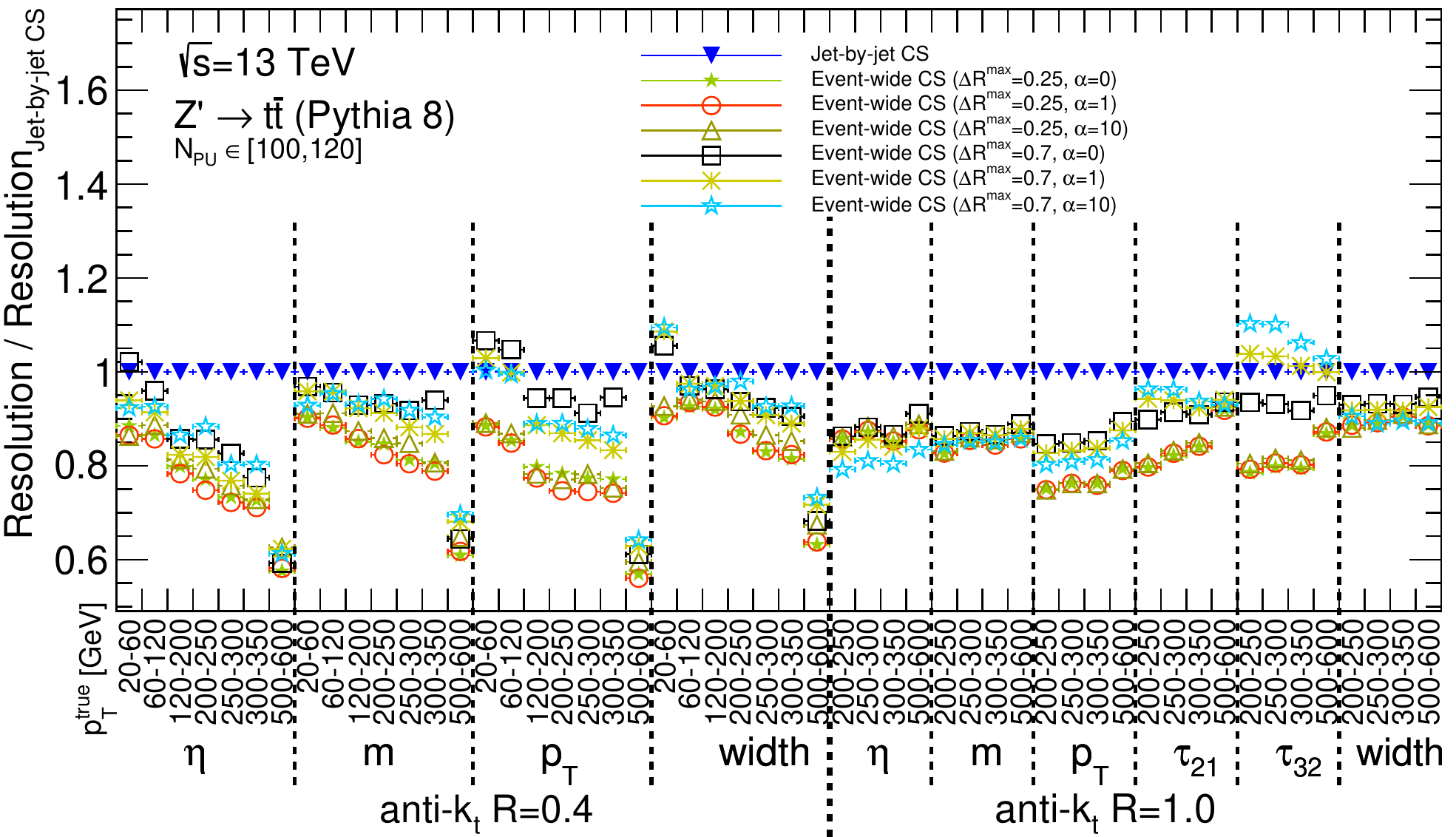}
  \caption{Dependence of resolution on the $\alpha$ parameter for the Event-wide CS.}
  \label{fig:alpha_resolution}
\end{figure}

\subsection{Ghost area \Aghost}
\label{app:ghost_area}
The smaller the value of \Aghost, the more densely are the ghosts distributed in the \yphi space which leads to better performance of the CS procedure. On the other 
hand, the smaller the value of \Aghost, the higher is the number of ghosts which implies a longer computational time. Therefore in practice, a compromise needs to be 
done. The optimal value of \Aghost may depend on the jet definition, granularity of the detector, and the \DeltaRmax parameter in the CS procedure. We found that 
$\Aghost=0.0025$ gives the best performance and using smaller \Aghost than 0.0025, does not lead to any significant improvement, it just brings a longer computational 
time. By using $\Aghost=0.01$, we observe on average 4-times faster correction with subtle worsening of performance (relatively up to ${\sim}2\%$ worse resolution for 
some observables).

\section{Choice of parameters for the ICS method}
\label{app:ICS_parameters}
We investigated the optimization of configuration of the ICS algorithm by varying: 
number of iterations, CS parameters (\DeltaRmax, $\alpha$, \Aghost) for each 
iteration, and the usage of ghost removal option. The performance studies presented here use the setup described in \secref{performance_setup}. We found that the best 
performance for the majority of studied observables and jet \pt ranges can be achieved with two iterations using ghost removal and parameters $\alpha=1$, 
$\Aghost=0.0025$, while the \DeltaRmax parameter is different for each iteration and also depends on the jet definition. For \aktfour jets, parameters 
$\DeltaRmax_1=0.2$ and $\DeltaRmax_2=0.1$ are found to be optimal for the first and second iteration, respectively. For \aktten jets, the parameters 
$\DeltaRmax_1=0.2$ and $\DeltaRmax_2=0.35$ are found to be optimal. We would like to emphasize, that these conclusions may not be perfect for a specific 
detector environment and/or for the case of heavy-ion collisions. Therefore, the experiments are encouraged to optimize the ICS configuration in their own environment.

In the following we provide a justification for the choice of the optimal ICS configuration based on performance studies. First we discuss ICS with two iterations 
and then we focus on the performance when using three or more iterations. The discussion of the choice of CS parameters \DeltaRmax, $\alpha$, $\Aghost$, which is provided in \appref{CS_parameters}, 
applies also for the ICS method. We use ghost area $\Aghost=0.0025$ and $\alpha=1$ in the following, since we have not observed any significant improvement by using 
smaller \Aghost or different $\alpha$ parameter. Therefore, we discuss just the parameter \DeltaRmax, the option of ghost removal, and the number of iterations in the 
following.

\subsection{ICS with two iterations}
  In general, it is expected that the ghost removal should help for cases when $\DeltaRmax_1\geq\DeltaRmax_2$, since then the ghost removal can guarantee that all the 
expected background deposition is subtracted (in case of not using ghost removal, ghosts, which were not fully subtracted in the first iteration, have no chance to be 
subtracted in the second iteration). For cases when $\DeltaRmax_1<\DeltaRmax_2$, all the expected background can be subtracted independently of the usage of the ghost 
removal option. The impact of the ghost removal may depend on properties of the background. In case when $\rho$ is correlated in close-by regions in the \yphi space, 
then it may be beneficial to use the ghost removal. For example, when one ghost is not fully subtracted in the first iteration, it means that the real $\rho$ in the 
vicinity of this ghost is smaller than it was originally estimated. Therefore when the $\rho$ is correlated in close-by regions, then the $\rho$ within larger area 
around this ghost should be also smaller than the estimated $\rho$ so it may be helpful to remove such ghosts. On the other hand, when the $\rho$ in close-by regions is 
anti-correlated, then it makes sense not to use the ghost removal.

\begin{figure}[!h]
  \centering
  \includegraphics[width=\textwidth]{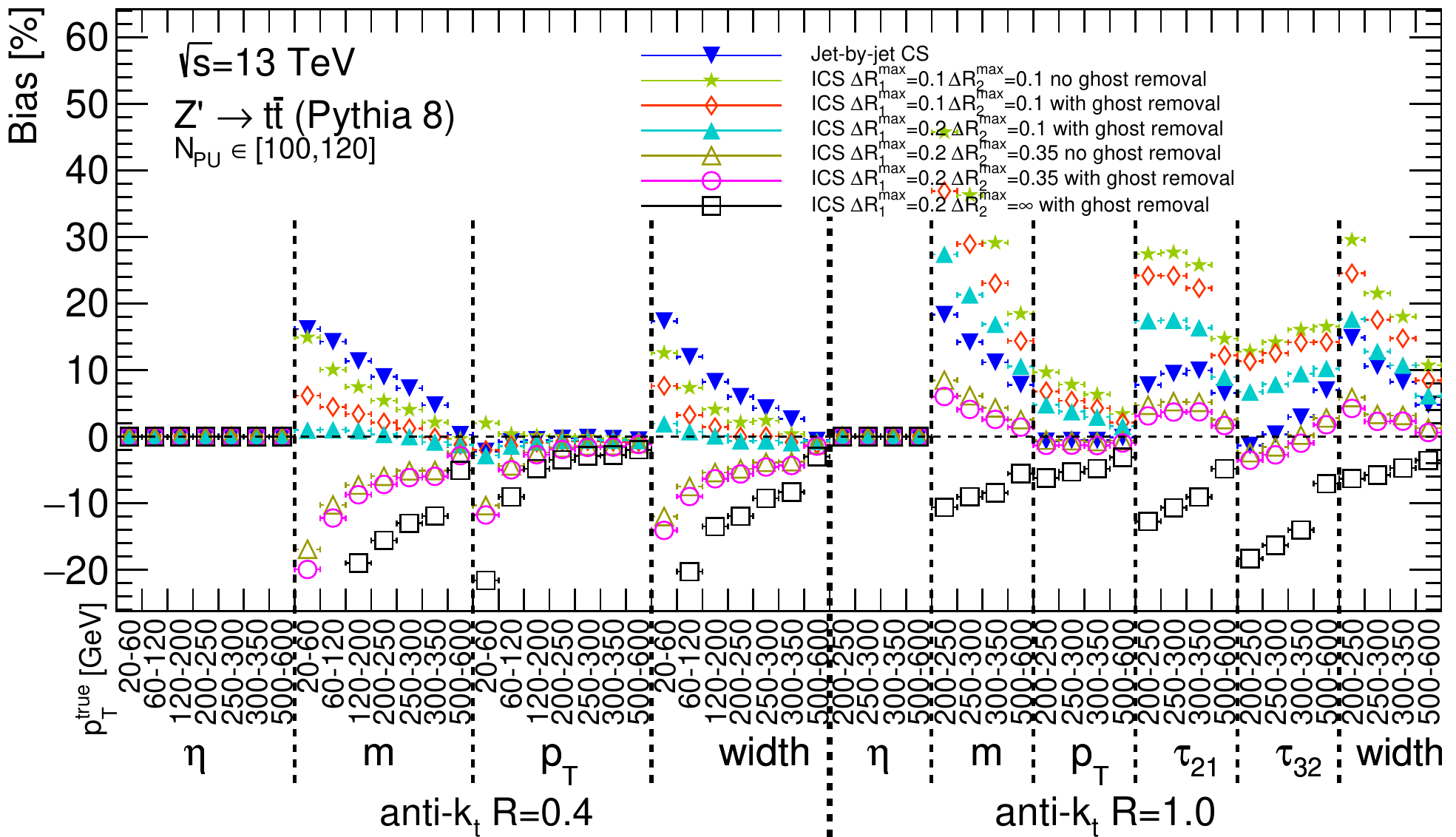}
  \caption{Bias for a representative set of configurations for the ICS method with two iterations. The \DeltaRmax parameter for the individual iterations is varied and the ghost removal option is used or not. The other CS parameters are set to $\alpha=1$ and $\Aghost=0.0025$ for each iteration.}
  \label{fig:ICSscan_2iter_bias}
\end{figure}

We investigated the performance of ICS with two iterations for reasonable values of \DeltaRmax parameter taking into account the findings in \appref{DeltaRmax}. For 
this, we used 50 configurations which were obtained by using \DeltaRmax parameter for the first iteration from the set $\{0.1,0.15,0.2,0.35,0.7\}$ and for the second 
iteration from the set $\{0.1,0.15,0.2,0.35,\infty\}$, and using or not using ghost removal. The bias and resolution for these 50 configurations vary significantly for 
the studied jet observables and jet \pt ranges. For each jet observable and jet \pt range, there is configuration which have maximal resolution at the expense of having 
higher bias compared to other configurations. To define the optimal configuration, we rather focus on minimizing the bias. Among the configurations which have \pt bias 
less than $3\%$ and bias for other observables less than $8\%$, we define the optimal configuration for each jet definition as the one which has maximal resolution for 
the highest number of jet observables and jet \pt bins. Based on this definition, we found that using the ghost removal option together with parameters 
$\DeltaRmax_1=0.2$ and $\DeltaRmax_2=0.1$ for \aktfour jets, and parameters $\DeltaRmax_1=0.2$ and $\DeltaRmax_2=0.35$ for \aktten jets provides the optimal 
configuration. We show the jet performance for a representative set of ICS configurations in \figsref{ICSscan_2iter_bias}{ICSscan_2iter_resolution}.

\begin{figure}[!h]
  \centering
  \includegraphics[width=\textwidth]{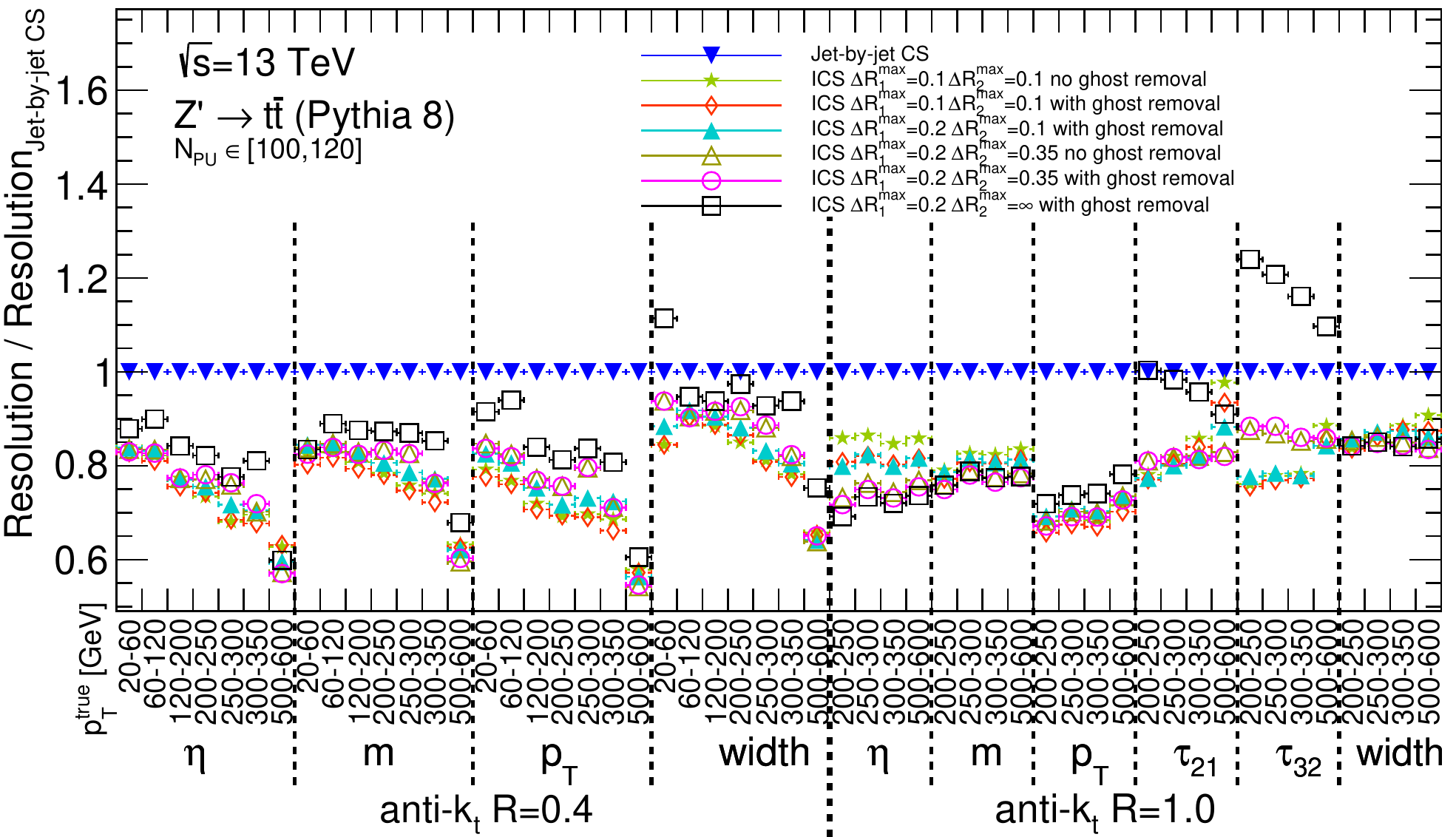}
  \caption{Resolution for a representative set of configurations for the ICS method with two iterations. The \DeltaRmax parameter for the individual iterations is varied and the ghost removal option is used or not. The other CS parameters are set to $\alpha=1$ and $\Aghost=0.0025$ for each iteration.}
  \label{fig:ICSscan_2iter_resolution}
\end{figure}

First we discuss the performance on \aktfour jets. The two shown configurations with $\DeltaRmax_1=0.1$ and $\DeltaRmax_2=0.1$ have the best resolution, but high bias 
for jet mass and jet width, especially for low-\pt ranges. We observe that the usage of the ghost removal option reduces the bias while the resolution stays similar for 
these \DeltaRmax values. The bias for jet mass and jet width can be further reduced by using ghost removal with $\DeltaRmax_1=0.2$ and $\DeltaRmax_2=0.1$ which is the 
configuration that we identified as the optimal one. By using higher $\DeltaRmax_2$, the bias and resolution for all jet observables get worse.

To get the optimal performance for \aktten jets higher \DeltaRmax values are preferred than for \aktfour jets, which is consistent with findings in \appref{DeltaRmax}. 
The configuration with $\DeltaRmax_1=0.2$ and $\DeltaRmax_2=0.35$ is found to be optimal. The ghost removal then further slightly improves the performance. The effect 
of ghost removal is much smaller for these \DeltaRmax parameters compared to values $\DeltaRmax_1=0.1$ and $\DeltaRmax_2=0.1$, which is expected since in the first 
iteration larger fraction of ghost \pt is subtracted so less scalar \pt is redistributed in the second iteration. The configuration $\DeltaRmax_1=0.2$ and 
$\DeltaRmax_2=\infty$ ensures that all expected background is subtracted, however it still seems that at wrong places, since the performance is worse for this 
configuration. At the same time the performance is still better compared to the performance of simple Event-wide CS with $\DeltaRmax=\infty$.

\subsection{ICS with three or more iterations}
\begin{figure}[!h]
  \centering
  \includegraphics[width=\textwidth]{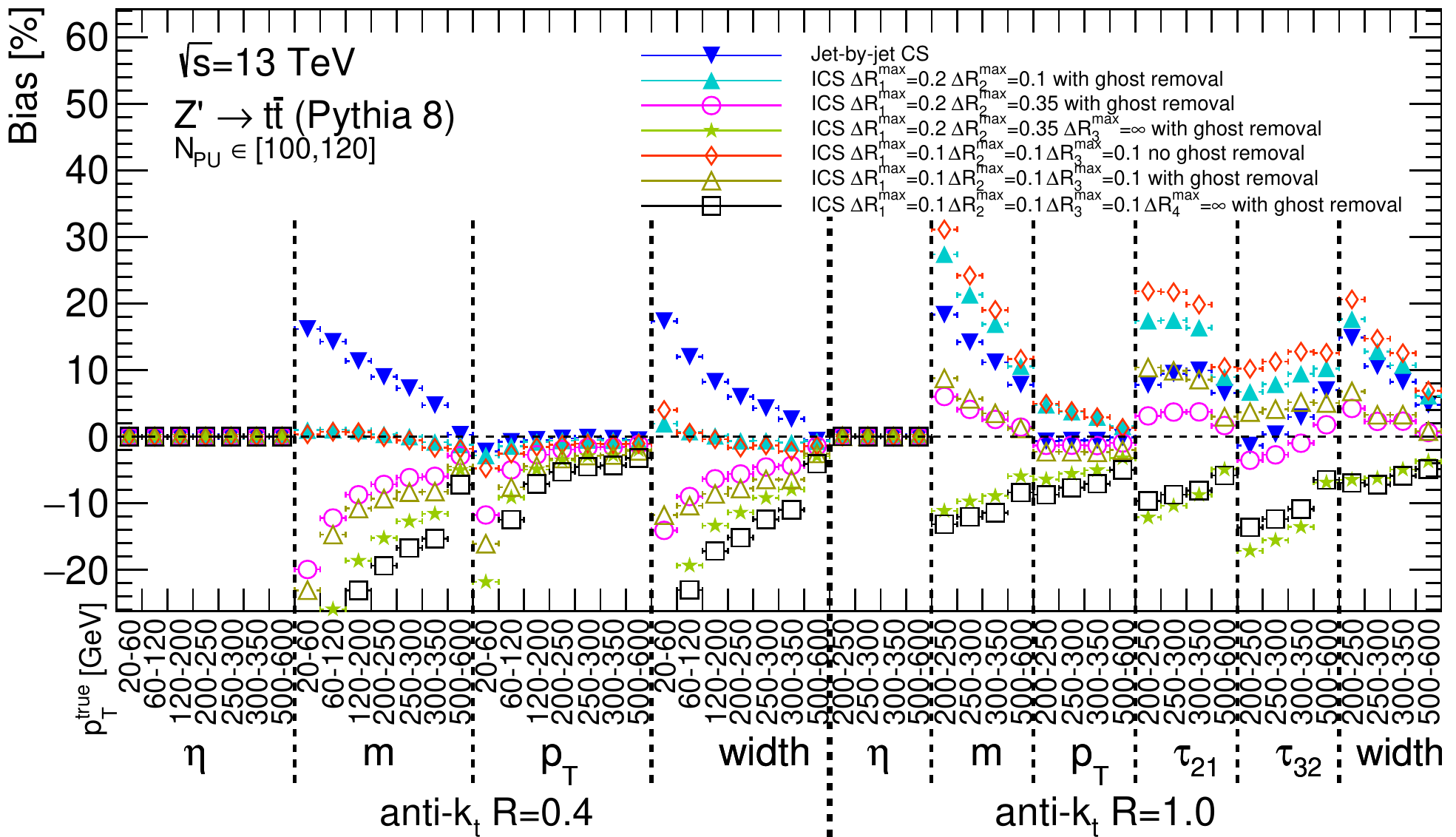}
  \caption{Bias for a representative set of configurations for the ICS method with three or more iterations. The \DeltaRmax parameter for the individual iterations is varied and the ghost removal option is used or not. The other CS parameters are set to $\alpha=1$ and $\Aghost=0.0025$ for each iteration.}
  \label{fig:ICSscan_3iter_bias}
\end{figure}

  We tried several configurations with three or more iterations. Based on the findings in the previous subsection, we used $\DeltaRmax_1$ and $\DeltaRmax_2$ values 
which exhibited only small overcorrection of jet \pt as a starting point. The jet performance for a representative set of configurations is shown in 
\figsref{ICSscan_3iter_bias}{ICSscan_3iter_resolution}. We did not find any significant improvement when using three or more iterations. However, some jet observables 
can have by ${\sim}10\%$ better resolution at the expense of slightly worse bias when using more iterations. This holds especially for the configuration with three 
iterations using $\DeltaRmax_1=0.1$, $\DeltaRmax_2=0.1$, and $\DeltaRmax_3=0.1$. Similarly as in the previous section, we observe that forcing to subtract all expected 
background by setting $\DeltaRmax_3=\infty$ is worse than keeping $\DeltaRmax_3$ finite (see e.g. configuration with $\DeltaRmax_1=0.2$, $\DeltaRmax_2=0.35$, and 
$\DeltaRmax_3=\infty$).

\begin{figure}[!h]
  \centering
  \includegraphics[width=\textwidth]{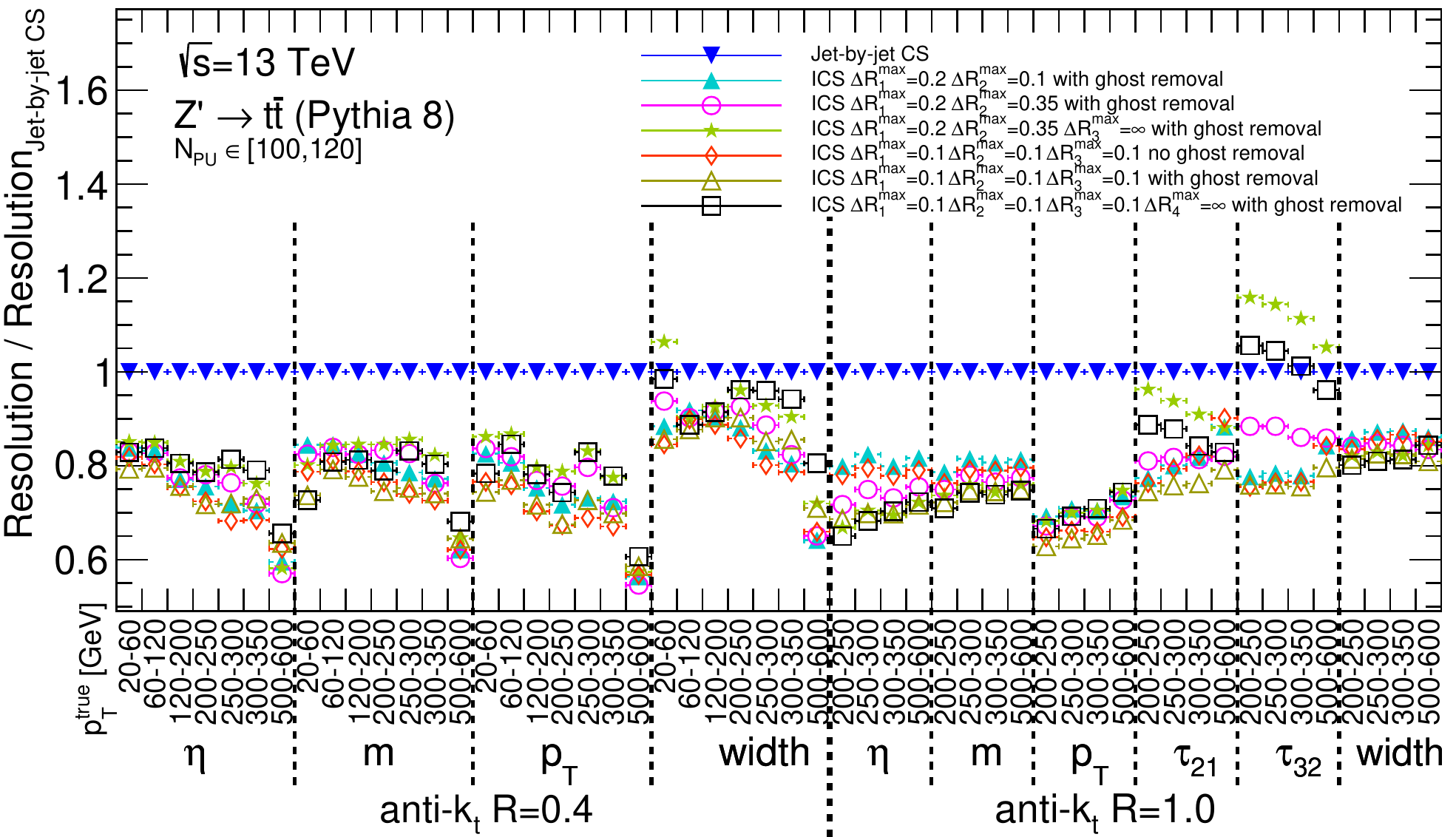}
  \caption{Resolution for a representative set of configurations for the ICS method with three or more iterations. The \DeltaRmax parameter for the individual iterations is varied and the ghost removal option is used or not. The other CS parameters are set to $\alpha=1$ and $\Aghost=0.0025$ for each iteration.}
  \label{fig:ICSscan_3iter_resolution}
\end{figure}

In this appendix, we identified and discussed an optimal configuration for which ICS preforms better than other background mitigation methods (see \secref{performance}) and 
provided a basic insight into the mechanisms that drives the performance. At the same time, the presented optimal configuration should not be viewed as the best 
configuration for all experiments and all backgrounds. The presented configuration should be viewed only as an advice which can be used as a starting point for further 
optimization within realistic environment of experiments.

\section{Treatment of massive particles}
\label{app:massive_inputs}
In the case of massless particles, there are three degrees of freedom for each particle \fourmomentum. The CS procedure corrects \pt of each particle, while the 
rapidity and azimuth are kept unchanged. In that way, it is ensured that the \akt jet clustering algorithm clusters the same particles\footnote{Up to small 
back-reaction biases coming from non-perfect correction which will be neglected in the following.}. Since $y=\eta$ for massless particles, also the momentum direction is 
kept unchanged which ensures good reconstruction of some jet observables such as the jet mass.

In the case of massive particles, there are four degrees of freedom for each particle \fourmomentum. The CS procedure corrects the \pt of each particle to compensate 
for the presence of background particles on average. Then the particle azimuth should be kept unchanged since there is no reason why the presence of background would 
change the azimuth. However, it is non-trivial to decide how to treat the remaining two degrees of freedom for the particle \fourmomentum. One straightforward option is to 
keep the mass of each particle unchanged. For the remaining degree of freedom, two obvious options are available: keep rapidity or pseudo-rapidity unchanged. Due to the 
fact that the \akt jet clustering algorithm uses difference of rapidities in the definition of the distance measure, there are the following caveats for the two 
options:
  \begin{enumerate}
  \item Keeping mass and rapidity unchanged -- in this case, it is ensured that the difference of rapidities in the \akt algorithm of each particle pair is not modified after the correction, so the particle content of the final jets can be similar to the particle content of true jets. However, in the limit when the corrected particle $\pt\rightarrow 0$, the $z$-component of momentum $p_z\rightarrow m\sinh{y}$ and pseudo-rapidity $\eta\rightarrow \pm\infty$, which means that after such correction, the central \akt jets may contain particles with non-negligible momentum pointing to totally different direction than the high \pt particles in that jet. This causes a very large bias for some jet observables such as jet mass.
  \item Keeping mass and pseudo-rapidity unchanged -- in this case, the difference of rapidities in the \akt algorithm of each particle pair can be modified which may lead to jets with different particle content compared to the true jets. Especially, in the limit when the corrected particle  $\pt\rightarrow 0$, the rapidity $y\rightarrow 0$ and the energy $E\rightarrow m$. Therefore after the correction, jets near $\eta=0$ typically contain a lot of constituents with non-zero mass but very small energy.
  \end{enumerate}

To summarize these two options we can say that although the particle mass is usually negligible compared to the particle energy, we see that it cannot stay unchanged due to the fact that the \akt jet clustering algorithm uses rapidity difference in its distance measure. Hence, it is important to modify the mass of the particles. There are several other options 
how to treat the two remaining degrees of freedom:
  \begin{enumerate}
  \setcounter{enumi}{2}
  \item Set the particle masses to zero. Keep the rapidity unchanged.
  \item Set the particle masses to zero. Keep the pseudo-rapidity unchanged.
  \item Do correction of variable $\mdelta=\sqrt{\pt^2+m^2}-\pt$ in the same way as it is described in the original CS procedure \cite{Berta:2014eza} which was based on the approach in \bibref{Soyez:2012hv}. Keep the rapidity unchanged.
  \item Do the same correction of \mdelta as in the previous option, just keep the pseudo-rapidity instead of rapidity unchanged.
  \item Keep the rapidity and pseudo-rapidity unchanged. This option is equivalent with scaling the original \fourmomentum by a factor corresponding to the ratio between the corrected and original \pt of the particle.
  \end{enumerate}

We note that by setting the particle mass to zero (options 3 and 4), a bias is introduced for $\npu=0$ by definition (since all the hard-scatter particles are modified 
to be massless). There is no such bias for options $5-7$. The options 5 and 6, which use the \mdelta correction, can still suffer from the same effects as described 
above the for options 1 and 2, respectively, thought with smaller impact on the performance. The options 3, 4, and 7 are not affected by the effects described above 
for options 1 and 2.

We investigated the performance of the individual options using the ICS method in the same conditions as described in \secref{performance_setup} but without the detector simulation. For these studies, we used ICS from \texttt{FastJet Contrib} version 1.042, where all the options for the treatment of massive particles are implemented. We found that the jet performance for options 1 and 2 is much worse than for options $3-7$, especially for the jet mass. The performance for jet $\eta$, jet \pt, jet width and subjettiness ratios is very similar among the options $3-7$. From the tested jet observables, only the jet mass has large dependence on the chosen strategy for correcting the particle \fourmomenta, therefore we discuss the performance for the jet mass in the following.

\begin{figure}[!h]
  \centering
  \begin{subfigure}{0.47\textwidth}
    \includegraphics[width=\textwidth]{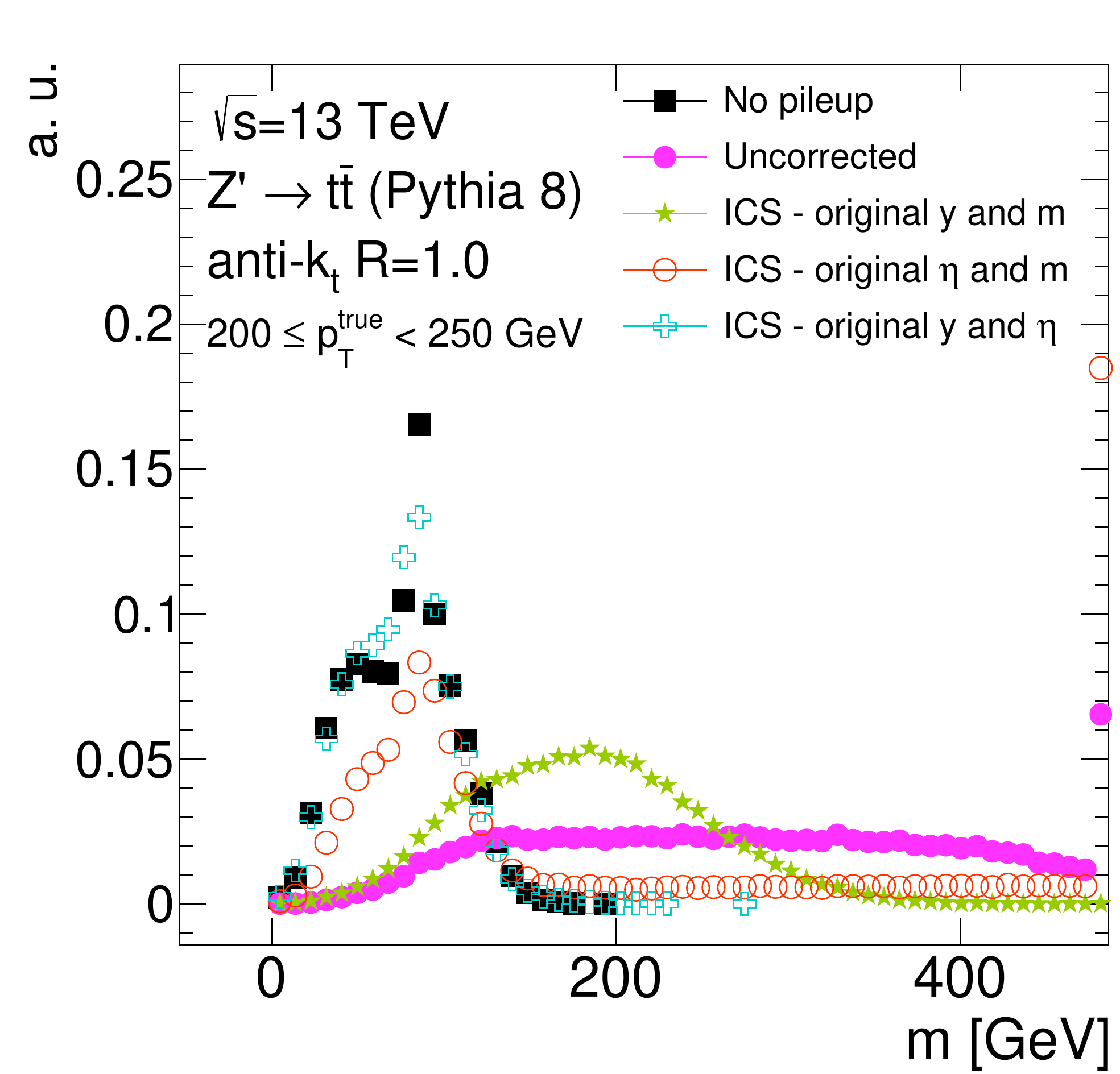}
    \caption{}
    \label{fig:Zprime:massiveParticles_originalMass}
  \end{subfigure}
  \begin{subfigure}{0.47\textwidth}
    \includegraphics[width=\textwidth]{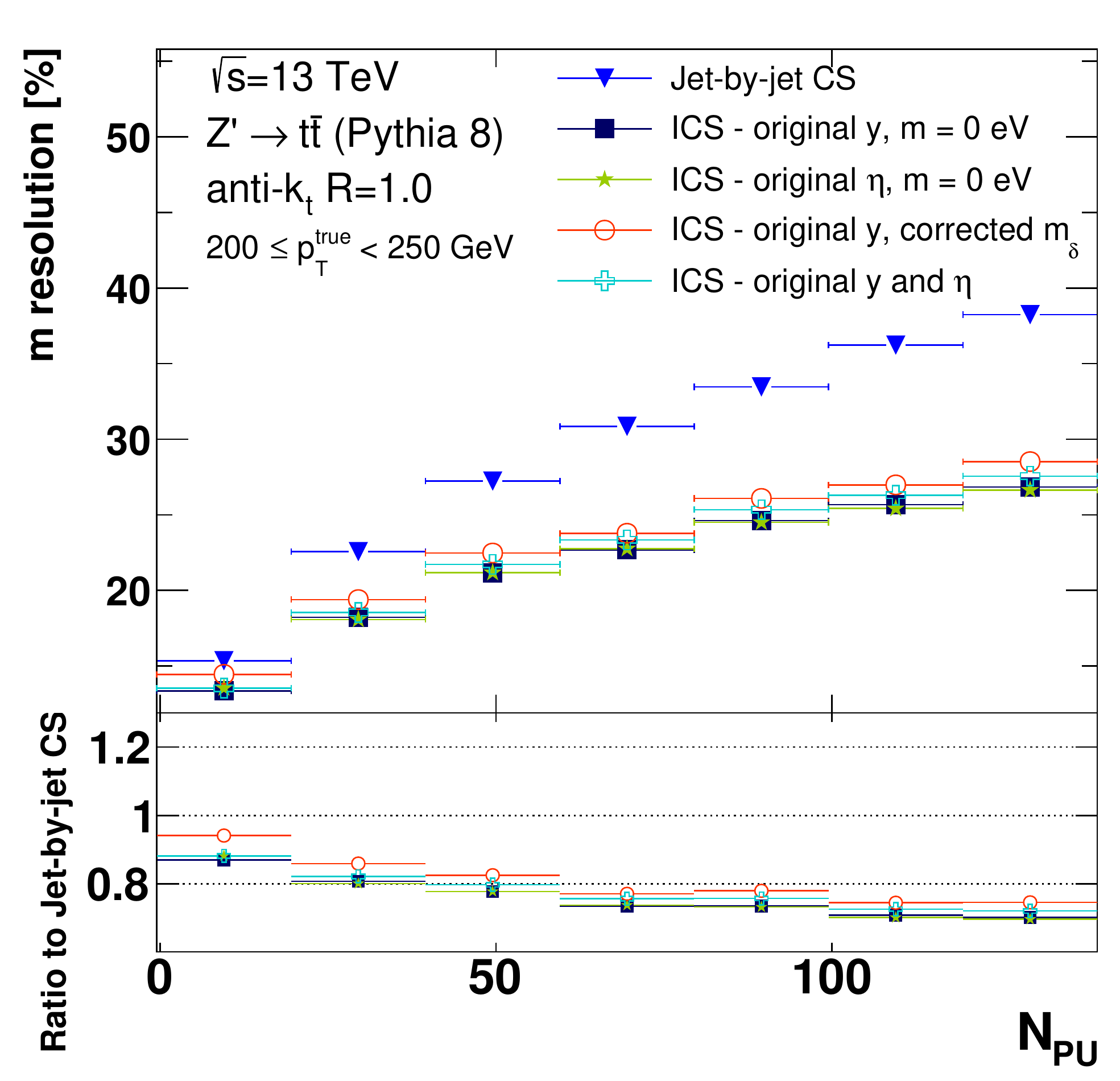}
    \caption{}
    \label{fig:Zprime:massiveParticles_correctedMass}
  \end{subfigure}
  \caption{Jet mass distribution (left) for the true jets, jets with pileup, and ICS corrected jets for three different approaches for the treatment of massive inputs: two approaches in which the particle mass is unchanged and one approach in which the rapidity and pseudo-rapidity is kept unchanged. Jet mass resolution (right) for six correction methods. The last bin is the overflow bin.}
  \label{fig:Zprime:massiveParticles}
\end{figure}

First, the importance of modifying the particle mass is demonstrated in \figref{Zprime:massiveParticles_originalMass} showing the jet mass distribution for \aktten jets 
with true $\pt = 200 - 250 \gev$ (containing mainly decay products from the $W$ boson). The true distribution is peaked around the $W$ boson mass. With \pileup, the 
original distribution is broadened and shifted to higher values. After ICS correction with keeping the original mass and rapidity of particles (option 1), the jet mass distribution is partially corrected, but still it is significantly shifted away from the true distribution. When using ICS correction with keeping the original mass and pseudo-rapidity (option 2), the corrected distribution agrees with the true distribution a bit better, but still it has a significant tail from jets biased towards high 
jet mass. In contrast, when using one of the options without keeping the original mass, e.g. option 7 (keeping original rapidity and pseudo-rapidity), the corrected 
mass distribution agrees much better with the true distribution.

Next, we investigated the bias and resolution of the jet mass systematically for options $3-7$. We found that option 6 has bias above $20\%$ and much worse resolution than the other four options. The option 5 has bias up to $10\%$ and resolution comparable with the options 3, 4, and 7. These remaining options have bias below $2\%$. The jet mass resolution for all these options using ICS as a function of $\npu$ is shown in \figref{Zprime:massiveParticles_correctedMass} compared to jet-by-jet CS (which uses option 3). As demonstrated in this figure, we found that setting the particle mass to zero (options 3 and 4) performs the best for the \aktten jets with true $\pt>200\gev$, while option 7 has only slightly worse resolution and option 5 has worse resolution by $5-10\%$. We found that the bias introduced by setting the particle mass to zero (options 3 and 4) gets negligible for $\npu>5$. We also observed that this jet mass bias is larger for low \pt \aktfour jets where the option 7 performs slightly better in general compared to options 3 and 4. 

We conclude from our studies, that keeping the original rapidity and pseudo-rapidity (option 7) leads to the optimal performance for the two studied jet definitions 
over the given phase space. However, the experiments using massive inputs are encouraged to check our findings for the exact setup of the analysis they intend to perform.

There is one other aspect of the background subtraction with massive particles. As mentioned in \secref{algo}, a user-defined parameter \maxEta is used, which 
ensures that only particles with $|\eta|<\maxEta$ are corrected by constructing the ghosts up to the same $|\eta|$ limit. However, the available software for $\rho$ 
estimation in \texttt{FastJet} uses particle rapidities and not pseudo-rapidities. Therefore, in the case of massive particles, it is important that the user uses 
only particles with $|\eta|<\maxEta$ to estimate the $\rho$ and also to derive the rapidity dependence used for background rescaling.

\acknowledgments


This project has received funding from the European Union's Horizon 2020 research and innovation programme under the Marie Sklodowska-Curie grant agreement No 797520. The work of MS was supported by Grant Agency of the Czech Republic under Grant 18-12859Y and by Charles University grants UNCE/SCI/013 and Progres Q47. DWM is supported by the National Science Foundation under Grant No. PHY-1454815.



\bibliography{ms}
\bibliographystyle{unsrt}

%
%
%
%




\end{document}